\documentclass[journal=nalefd,manuscript=article,preprint]{achemso}

\setkeys{acs}{etalmode = truncate, maxauthors=10}

\usepackage[colorlinks=true, citecolor=black, linkcolor=black, urlcolor=black]{hyperref}

\usepackage[all]{hypcap}
\usepackage{amsmath}
\usepackage{enumerate,bm}
\usepackage{graphicx}
\usepackage{grffile} 
\usepackage{color}
\usepackage{esint}
\usepackage{textpos}
\usepackage[usenames,dvipsnames]{xcolor}
\usepackage{subfigure}
\usepackage{flushend}
\usepackage{tabularx}
\usepackage{amssymb}
\usepackage{float}
\usepackage{subfigure}
\usepackage{setspace}
\usepackage{pdfpages}

\newcommand{\hamburg}{Max Planck Institute for the Structure and Dynamics of Matter, Luruper Chaussee 149, 22761 Hamburg, Germany.

Center for Free-Electron Laser Science and Department of Physics, University of Hamburg, Luruper Chaussee 149, 22761 Hamburg, Germany
}

\newcommand{\ccq}{Center for Computational Quantum Physics (CCQ), The Flatiron Institute, 162 Fifth avenue, New York NY 10010.}

\newcommand{\unipa}{Dipartimento di Fisica e Chimica, Universit\'a degli Studi di Palermo, Via Archirafi 36, I-90123, Palermo, Italy}

\title{Cavity control of Excitons in two dimensional Materials}

\date{\today}

\author{Simone~Latini}
\email{simone.latini@mpsd.mpg.de}
\affiliation{\hamburg}
\altaffiliation{Contributed equally to this work}

\author{Enrico~Ronca}
\email{enrico.ronca@mpsd.mpg.de}
\affiliation{\hamburg}
\altaffiliation{Contributed equally to this work}

\author{Umberto~De~Giovannini}
\affiliation{\hamburg}
\alsoaffiliation{\unipa}

\author{Hannes~H\"ubener}
\affiliation{\hamburg}

\author{Angel~Rubio}
\email{angel.rubio@mpsd.mpg.de}
\affiliation{\hamburg}
\alsoaffiliation{\ccq}

\begin{document}

\begin{abstract}
We propose a robust and efficient way of controlling the optical spectra of two-dimensional materials and van der Waals heterostructures by quantum cavity embedding. 
The cavity light-matter coupling leads to the formation of exciton-polaritons, a superposition of photons and excitons. Our first principles study demonstrates a reordering and mixing of bright and dark excitons spectral features and in the case of a type II van-der-Waals heterostructure an inversion of intra and interlayer excitonic resonances. We further show that the cavity  light-matter coupling strongly depends on the dielectric environment and can be controlled by encapsulating the active 2D crystal in another dielectric material.
Our theoretical calculations are based on a newly developed non-perturbative many-body framework to solve the coupled electron-photon Schr\"odinger equation in a quantum-electrodynamical extension of the Bethe-Salpeter approach. This approach enables the ab-initio simulations of exciton-polariton states and their dispersion from weak to strong cavity light-matter coupling regimes. Our method is then extended to treat van der Waals heterostructures and encapsulated 2D materials using a simplified Mott-Wannier description of the excitons that can be applied to very large systems beyond reach for fully ab-initio approaches.
\end{abstract}
\maketitle


\paragraph*{Introduction}
Excitons dominate the optical properties of two-dimensional semiconductors. Single layers of Transition Metal Dichalcogenides (TMDs) have been under intense investigation for their excitonic properties \cite{ManzeliNatureRevMat2017,WangRevModPhys2018,UgedaNatureMater2014}. Their weak electronic screening \cite{RamasubramaniamPhysRevB2012,CudazzoPhysRevB2011}, a consequence of the reduced dimensionality, leads to the formation of strongly bound bright and dark excitons which play a fundamental role in a large variety of optoelectronic, spintronic and valleytronic properties \cite{SchaibleyNatureRevMat2016}. Furthermore, single layers of TMDs can be stacked in multilayer heterostructures allowing for device engineering with a high degree of freedom  \cite{KimScienceAdv2017,GaoNanoLett2012,GeimNature2013}. Among other designs, bilayers of TMDs with a type II band alignment enables the creation of interlayer excitons, bound electron-hole pairs where the charges are physically confined in two different layers, which show a great potential in photovoltaic applications \cite{HongUltrafast2014, JinUltrafast2018} 

Due to their strong coupling to electromagnetic radiation, TMD excitons represent ideal candidates to study strong light-matter coupling in optical cavities \cite{FlattenSciRep2016,ChervyACSPhot2018,SlootskyPhysRevLett2014,LiuNaturePhoton2014,SunNaturePhoton2017}. In an optical resonator, excitons interact with the quanta of light, generated by the spatial confinement of the cavity, resulting in the formation of new hybrid states with partial matter and partial light character, the \textit{exciton-polaritons} \cite{flick2017atoms, flick2018ab, ruggenthaler2018quantum}. These states are inherently different from the bare excitonic states in the material and therefore a variety of novel phenomena can be expected. 
Envisioned examples comprise new topological phases \cite{SheikhanPhysRevA2016}, light-induced superconductivity \cite{Sentef:2018gp}, exciton-polariton condensates \cite{ByrnesNaturePhys2014} and superradiance from  exciton insulators\cite{Mazza2019}.

\begin{figure}
\centering
\includegraphics[width=0.6\columnwidth]{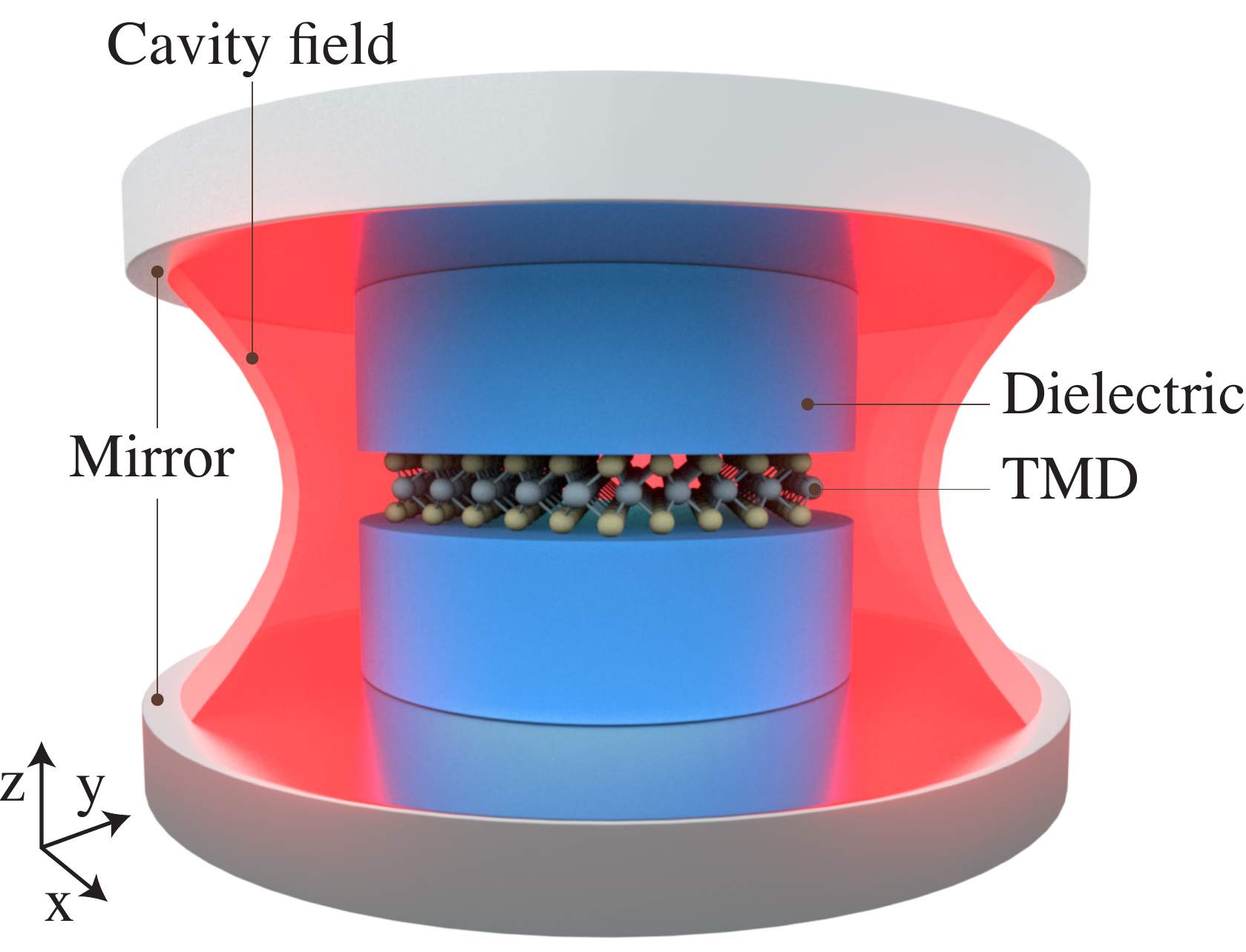}
   \caption{\label{fig:fig1}
   Schematic representation of an encapsulated monolayer TMD in an optical resonator (cavity) including the dielectric media that sustain the 2D material.
    } 
\end{figure}

Here, we present a first-principles theoretical framework, referred to as \\ quantum-electrodynamical Bethe-Salpeter Equation (QED-BSE), designed to describe the coupling of excitonic states in solid state materials embedded in a quantum optical cavity. Our approach goes beyond a simple model Hamiltonian description of exciton polariton by providing a quantitative description of exciton-polariton directly comparable to the experiments. The method requires the exact diagonalization of the exciton-photon equation obtained by extending the widely used many-body Bethe-Salpeter (BSE) formalism \cite{SalpeterPhysRev1951,RohlfingPhysRevLett1998,RohlfingPhysRevB2000,OnidaRevModPhys2002} to QED, under the approximation that the electron-electron interaction is not affected by the photon dressing. This method represents an alternative to the recently proposed quantum-electrodynamical density functional theory\cite{Tokatly:2013co,RuggenthalerPhysRevA2014,Pellegrini:2015kq,FlickProcNatAcSci2015,flick2017atoms,Rokaj:2018dn,flick2018ab,ruggenthaler2018quantum,DavisArXiv2018}, the latter describing the cavity mediated electron-photon interaction through effective exchange-correlation functionals of the electron and photon densities and currents. Our current method can be used as a starting point to determine such functionals as done in the past for describing the many-body excitonic properties within time-dependent (current) density functional theory\cite{OnidaRevModPhys2002}. In this work we demonstrate that by embedding a two-dimensional crystal in a cavity as sketched in Fig.~\ref{fig:fig1}, excitonic optical activity and energetic ordering can be controlled through cavity size, light-matter coupling strength and encapsulation in a dielectric material.
While the reordering of excitonic resonances has been qualitatively demonstrated in the case of molecular systems by means of model hamiltonians consisting of coupled oscillators for the photonic and excitonic fields\cite{DuChemSci2018, PolakArXiv2018}, here we aim to quantitatively demonstrate similar effects in the context of extended system.\\
Within dipolar selection rules there are bright excitons (BE), i.e. accessible as direct transition from the groundstate upon photon absorption, and dark excitons (DE) which are instead not accessible in linear optical spectroscopy \cite{CaoPhysRevLett2018}. The difficulty to access DE with standard optical spectroscopy poses a challenge for their detection and recent experimental and theoretical works have proposed sophisticated techniques to access them in TMDs \cite{YeNature2014,Loh:2017cw}.  Here we address and control the activity of symmetry-forbidden DE through the coupling of the TMD crystal with a quantum cavity which allows direct transitions from the groundstate. 
More complex types of excitons can be obtained if different monolayers of TMDs with a type II band alignment are stacked together in a van der Waals heterostructure (vdWH)\cite{LatiniInterlayer2017}. In particular
these type II heterostructures can host excitons where the electron and the hole are localized in different layers, {\it interlayer} excitons, in addition to excitons localized on a single layer, the {\it intralayer} ones. Here we show how the coupling to an optical cavity can be used to tune the order of the spectral resonances of the different types of excitons, providing a further knob for the design of optical devices based on vdWHs.  
We also show that the ab initio QED-BSE results can be described by an alternative approach based on the Mott-Wannier model, which we refer to as MW-QED, represents a computationally inexpensive method able to simplify the QED-BSE approach by simplifying the description of excitons\cite{GrossoBook2000,KeldyshJETPLett1979,LatiniPhysRevB2015}. The MW-QED accurately reproduces the results of the QED-BSE opening the way to the modeling of more complex systems, such as 2D heterostructures, Moir\'e patterned twisted bilayer systems and others, in optical cavities. 
Finally we use the MW-QED to characterize the effect produced by a dielectric on the polaritonic spectrum of a single layer of TMD.  

\paragraph*{Theoretical Framework}
The electronic structure of a strongly coupled light-matter system in a cavity requires a non-perturbative treatment of the electron-photon interaction. 
Fundamentally, light-matter coupling is described by quantum-electrodynamics where the many-body electron and photon state is described by a combined Hamiltonian. Specializing this Hamiltonian to the case of a 
a single cavity mode of frequency $\Omega$ and long-wavelength (dipolar) electron-photon coupling we obtain the first-principles Hamiltonian for perfect loss-less cavity in velocity gauge:
\begin{equation}\label{eq:QEDgen}
\hat{H}_{\rm{QED}} =  \hat{H}_{\rm{el}} + \Omega \hat{a}^\dagger \hat{a} +  N_{\rm{el}}\frac{A_0^2}{2}(\hat{a}^\dagger + \hat{a})^2+ 
  A_0 \sum_{ij\bf{k}}\left(\langle\phi_{i\bf{k}} | \hat{e}\cdot\hat{p} | \phi_{j\bf{k}}\rangle \hat{d}^\dagger_{i\bf{k}}\hat{d}_{j \bf{k}} \hat{a}^\dagger +  h.c.\right),
\end{equation}
where $a^\dagger$ and $a$ are the photon creation and annihilation operators respectively, $\hat{H}_{\rm{el}}$ the many-body electronic Hamiltonian, $\hat{d}^\dagger_{i\bf{k}},\hat{d}_{j \bf{k}}$ the electronic creation and annihilation operators (with $i,j$ band indices and $\bf{k}$ wavevectors in the first Brillouin zone), $\hat{p}$ the single particle momentum operator, $N_{\rm{el}}$ the number of electrons, $\hat{e}$ the polarization of the photon and $A_0$ is the vector potential amplitude.
Since the coupling of photons to the electronic structure occurs via the creation/annihilation of neutral electron-hole pairs, it is natural to approximate the eigenstates of the many-body electronic Hamiltonian by its excitonic eigenstates, i.e. $\hat{H}_{\rm{el}}|\Psi_{n}^{\rm{exc}}\rangle\simeq\epsilon_n^{\rm{exc}}|\Psi_{n}^{\rm{exc}}\rangle$. Within this assumption we can accurately describe a regime where the concentration of excitons is low enough to neglect exciton-exciton interaction, which is the case for non-pumped cavities. Under the approximation that the electron-hole Coulomb interaction that binds the exciton is not affected by the photons in the cavity, i.e. the screened Coulomb interaction is not dressed by the cavity photons, the QED Hamiltonian can be expanded in the excitonic basis as follows:
\begin{equation}\label{eq:QED}
\langle\Psi_{n}^{\rm{exc}}|\hat{H}_{\rm{QED}}|\Psi_{m}^{\rm{exc}}\rangle =
\left[\epsilon_{n}^{\rm{exc}} + \Omega \hat{a}^\dagger \hat{a}  + N_{el} \frac{A_0^2}{2}(\hat{a}^\dagger + \hat{a})^2\right]\delta_{nm} +
A_0 \left(\mathcal{M}^{\rm exc}_{nm} \hat{a}^\dagger + \mathcal{M}^{\rm exc *}_{mn} \hat{a}\right), 
\end{equation}
where $\mathcal{M}^{\rm exc}_{nm} = \sum_{ij\bf{k}} \langle\phi_{i\bf{k}} | \hat{e}\cdot\hat{p} | \phi_{j\bf{k}}\rangle \langle\Psi_{n}^{\rm{exc}}|\hat{d}^\dagger_{i\bf{k}}\hat{d}_{j \bf{k}}|\Psi_{m}^{\rm{exc}}\rangle $
are excitonic matrix elements of the bilinear dipole electron-photon coupling and $\phi_{i\mathbf{k}}$ are single particle Bloch functions. 
In this framework the excitonic states can be expressed as a linear combination of singly excited electronic determinants, where the coefficients of the linear combination are given by the solution of the BSE \cite{RohlfingPhysRevLett1998,RohlfingPhysRevB2000,OnidaRevModPhys2002},
$
|\Psi_{n}^{\rm{exc}}\rangle = \sum_{cv\bf{k}} A_{cv\bf{k}}^{n} \hat{d}^\dagger_{c\bf{k}}\hat{d}_{v \bf{k}}|\Psi_0\rangle
$
with $A_{cv\bf{k}}^{n}$ the BSE coefficients, or envelope functions and $c$ and $v$ indices running over conduction and valence bands respectively. The electronic groundstate $|\Psi_0\rangle$ instead is assumed to be a single determinant of only valence states
and we define $|\Psi_{\rm{n}=0}^{\rm{exc}}\rangle=|\Psi_0\rangle$.

In this work we consider the non-dispersive (momentum independent) excitonic states localized around the K-points of the Brillouin zone of the TMD, which are the most relevant for the optoelectronic properties of TMDs. These excitonic states occur in two spin-orthogonal non-hydrogenic series, each of which is accurately described by a spin-independent two-band BSE, where only a single valence and a single conduction band are taken into account \cite{LatiniPhysRevB2015}. With this simplification the matrix elements appearing in Eq.~(\ref{eq:QED}) have one of the two following structures:
\begin{align}\label{eq:ME_GS}
\mathcal{M}^{\rm exc}_{0n} &=\sum_{\rm{\bf{k}}} A_{\rm{\bf{k}}}^{n}\langle \phi_{v\bf{k}} | \hat{e}\cdot\hat{p} | \phi_{c\bf{k}}\rangle\\
\mathcal{M}^{\rm exc}_{nm} &= N_{el}\sum_{\rm{\bf{k}}} A_{\rm{\bf{k}}}^{m*}A_{\rm{\bf{k}}}^{n}\left[\langle\phi_{c\bf{k}} | \hat{e}\cdot\hat{p} | \phi_{c\bf{k}}\rangle - \langle \phi_{v\bf{k}} | \hat{e}\cdot\hat{p} | \phi_{v\bf{k}}\rangle\right]\label{eq:ME_EXC}
\end{align}
The matrix elements in Eq.~(\ref{eq:ME_GS}) are those that dictate the dark/bright nature of the excitonic states. Those in Eq.~(\ref{eq:ME_EXC}), instead, are mixing the character of the excitonic states so that bright to dark or dark to bright transitions can occur. 

Below we show first principles results for MoS$_2$ obtained by first solving the BSE with the GPAW code \cite{MortensenPhysRevB2005,EnkovaaraJPhysCondMat2010} (see supplementary material for calculation details) and subsequently diagonalizing the QED Hamiltonian of Eq.~(\ref{eq:QED}) in a mixed exciton-photon product state basis $|\Psi^{\rm exc}_{n}\rangle\otimes|\gamma\rangle$, where $|\gamma\rangle$ are the eigenfunctions of the photonic harmonic oscillator. We then extract energies ($E^{\rm pol}_{I}(\Omega)$) and corresponding eigenfunctions ($|\Psi^{\rm pol}_I(\Omega)\rangle = \sum_{n \gamma} C^{I}_{n\gamma}|\Psi^{\rm exc}_{n}\rangle\otimes|\gamma\rangle$) for the exciton-polariton eigenstates. As for the photonic part, we consider the first non-zero photon mode ($\Omega=\pi c/L_{\perp}$) with out-of-plane wavevector  and in-plane electric field which is able to couple to the MoS$_2$ in-plane excitonic dipole. For our quasi-2D cavity configuration the vector potential amplitude $A_0$ is frequency independent, unlike in the 3D case, indeed $A_0=1/\sqrt{S L_{\perp} \Omega}=1/\sqrt{2\pi c S}$, where S is the in-plane area of the cavity. Furthermore we stress that the single mode approximation is valid for polaritons with energies lower than the energy of the second photon mode ($\Omega=2\pi c/L_{\perp}$). In the supplementary material we show that including higher energy modes does not change the conclusions discussed below. We note that diagonalization of Eq.~(\ref{eq:QED}) is equivalent to solving a CI-singles in the presence of a photon mode. In particular, excited states of the material are expressed as linear combinations of singly excited determinants starting from the ground state and subsequently coupled, through the diagonalization, to the photonic degrees of freedom.

\paragraph*{Results}
\begin{figure}
\centering
\includegraphics[width=1\columnwidth]{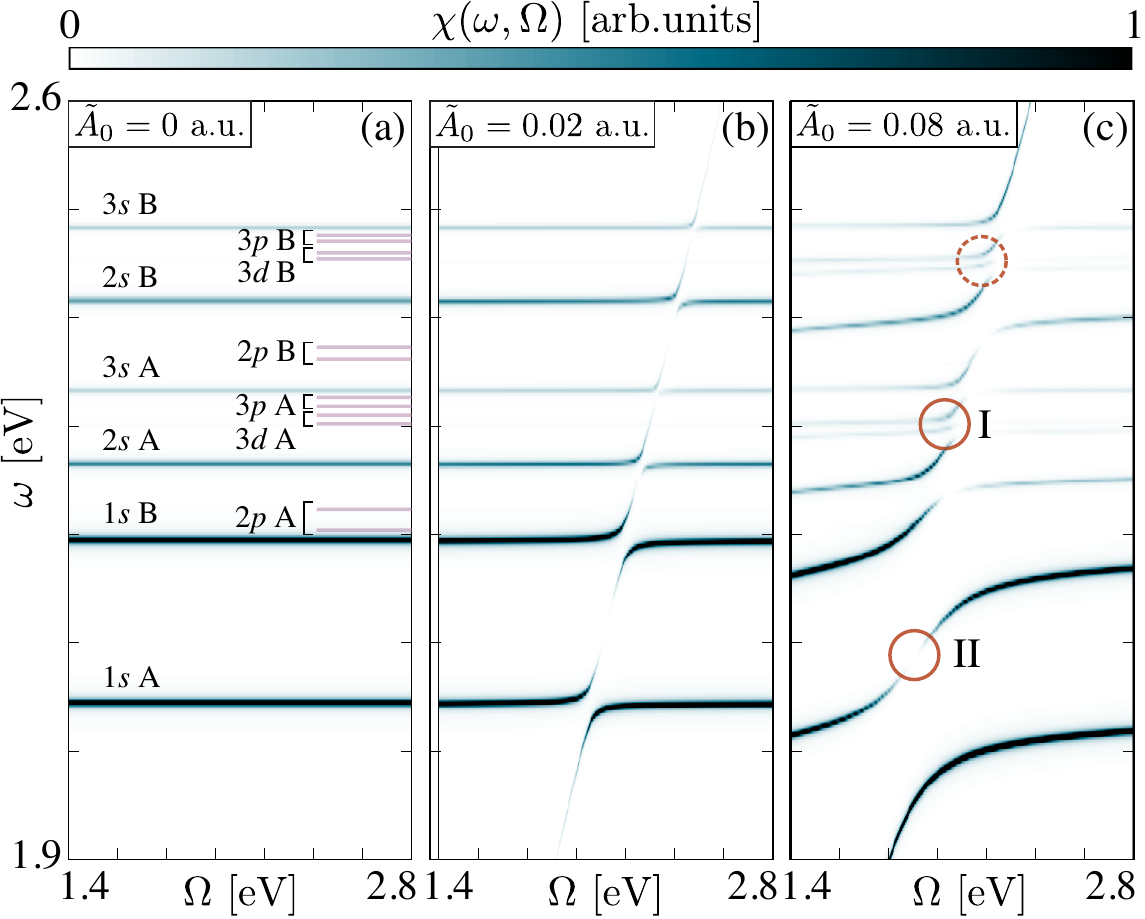}
   \caption{\label{fig:fig2}
   Exciton-polariton spectra of MoS$_2$ in a quantum cavity as a function of cavity mode frequency $\Omega$ and for three different coupling strength $\tilde{A}_0$ in a.u.. (a) For no coupling to the cavity mode the optical spectrum displays a series of spin-orbit split A and B excitons where only excitons with $s$ symmetry appear, i.e. are bright. DE are reported in red segments. (b) For a weaker coupling, excitons hybridize with the cavity photon to form exciton polaritons resulting in Rabi splitting of the bright exciton lines. (c) For the stronger coupling value, the Rabi splitting becomes more pronounced and additional polariton features appear where bright polariton branches cross dark exciton states. The broadening parameter used for all plots is $\eta=1.4$ meV.    
    } 
\end{figure}

In order to visualize the experimentally measurable exciton-polariton dispersion, we calculate the matter component of the full optical response $\chi(\omega,\Omega)$ applying linear response theory on the polaritonic states\cite{DavisArXiv2018,ruggenthaler2018quantum}:
\begin{equation}\label{eq:response}
    \chi(\omega,\Omega) = \sum_{I}\frac{\mathcal{M}^{\rm pol}_{0I}\mathcal{M}^{\rm pol}_{I0}}{\omega-E^{\rm pol}_{I}(\Omega)+E^{\rm pol}_{0}(\Omega)+i\eta}
\end{equation}
where $\mathcal{M}^{\rm pol}_{IJ}=\langle\Psi^{\rm pol}_I|\hat{p}|\Psi^{\rm pol}_{J}\rangle=\sum_{nm\gamma} C^{I*}_{n\gamma}C^{J}_{m\gamma}\mathcal{M}^{\rm exc}_{nm}$ and $i\eta$ is a small artificial imaginary broadening.
Such a quantity represents the independent particle polarizability of the polariton to a weak external field coupling with the matter component. An equivalent quantity can be also formulated to investigate the photonic counterpart of the polariton.
Comparison between the matter and photon spectral functions is presented in Fig.~S5 of the Supporting Information.

The formation of exciton-polaritons results in a richer optical response as compared to the bare excitonic response. In Fig.~2 we show the cavity polariton spectrum of MoS$_2$ as a function of mode energy and for different coupling strength $\tilde{A}_0=A_0/N_{\rm{el}}$. Here the coupling strength arises from a microscopical definition 
while in practice it can be controlled by the cavity geometry and manufacturing parameters \cite{FlickNanophotonics2018}.

Without any electron-photon coupling ($\tilde{A}_0=0$), the spectrum features only the bare BEs as shown in Fig.~2(a), characterized by the common spin-orbit split A and B exciton series (see supplementary material)\cite{qiu2013optical}. Bound excitons in TMDs have atom like spherical symmetries, following the usual pattern of $s,p,d$ angular momentum quantum numbers and consequently are subject to atomic like selection rules \cite{YeNature2014}. Only excitons with $s$-type symmetry are accessible from the ground state through direct absorption of photons. The $p$ and $d$-type excitons instead can be classified as DEs and are indicated with lines in Fig.~2(a) for reference. 

In the weaker coupling regime (c.f. Fig.~\ref{fig:fig2}(b)) the formation of an exciton-polariton is accompanied by the characteristic appearance of avoided crossing in the energy dispersion, which results from the hybridization between the linear dispersing photon branch and the non-dispersive exciton branches. For a given cavity frequency within the hybridisation region, this is detectable as a splitting of the exciton peak into two separate peaks -- the Rabi splitting. The formation of lower and upper polariton branches results in a reordering of the energy of bright and dark excitonic resonances which can be of technological relevance for stabilizing long-lived dark excitonic states.

For the stronger coupling (c.f. Fig.~2(c)), the interaction between the cavity photon and the DE results in the appearance of new spectral features marked by circles in the figure. Specifically, we can identify additional splittings (I)/interruptions (II) signature of a cavity induced optical modulation. 
It is important to mention that the mixing between bright and dark excitons discussed here is possible only for DE forbidden by symmetry of the wave function\cite{YeNature2014}. Different kind of dark excitons like the ones originating from spin-selection rules or beyond dipole selection rules are not discussed in this work\cite{MalicPhysRevMat2018}.
In this stronger coupling regime ($\tilde{A}_0=0.08$) we performed calculations also for different TMDs (MoSe$_2$, WS$_2$ and WSe$_2$, see supplementary material) observing qualitatively equivalent features. It is worth noting that for the most bound exciton of WS$_2$ we obtain a Rabi splitting of $\approx22$  meV in line with experimental values measured by Flatten et al. ($\approx20-70$ meV)\cite{FlattenSciRep2016}. While it is hard to asses the exact coupling conditions under which the experiment has been performed, this comparison shows that the coupling values used here are on the lower end of the experimentally accessible values. The coupling values used in the results are also in line with the typical sizes of the cavity used in the experiments, which are in the $\mu m$-$nm$ range. In particular, using the $\tilde{A}_0$ definition presented above, a coupling value of $\approx 0.08$ a.u. and an electron density of 0.114 el/\AA$^2$, like in the case of MoS$_2$,  corresponds to a cavity area of of about $0.25\mu m^2$.

\begin{figure}
\centering
\includegraphics[width=1.\columnwidth]{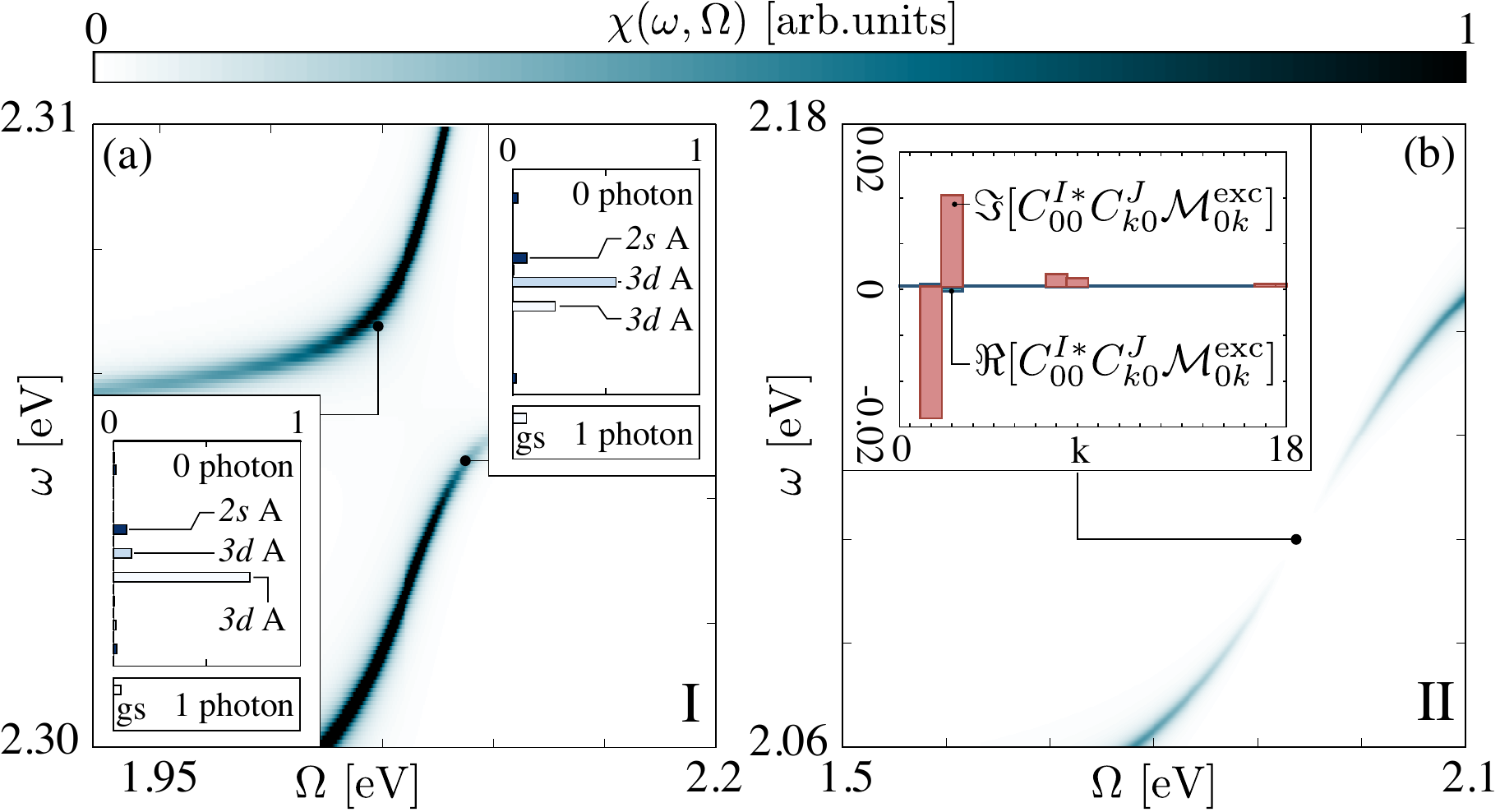}
   \caption{\label{fig:fig3}
   (a) Enlarged view of the dark exciton for the region highlighted in Fig.~\ref{fig:fig2} by I. The insets show the composition of the polariton eigenstates (see text) for points on the lower and upper branch. The compositions are taken as the expansion coefficients of the polaritonic eigenstates in exciton-photon product state basis ${|\Psi^{\rm exc}_{n}\rangle\otimes|\gamma\rangle}$. The involved excitonic states are categorized in boxes according to the photon sectors they are associated, i.e. $|n\rangle=|0\rangle, |1\rangle$. (b) Enlarged view of a point where the polariton vanished exactly between two consecutive bright exciton lines, highlighted in Fig.~\ref{fig:fig2} by II. The inset shows the real and imaginary parts of the matrix elements contributing to this feature (see text).} 
\end{figure}

Further insight into the structure of these dark polaritons is provided by the analysis shown in Fig.~\ref{fig:fig3}. The panels quantify the contribution of the different excitonic states to the optically active polariton states at specific cavity energy along the dispersion branches shown in Fig.~2. The decomposition of the polariton in Fig.~\ref{fig:fig3}(a), which is a zoom-in of the circle (I) in Fig.~\ref{fig:fig2}(c), shows that it is predominantly the $3d$ DE that is creating the polaritonic eigenstate. However, it is only because of the photon mediated coupling to the $2s$ BE that the $3d$ DE acquires a finite optical cross section. More specifically the coupling between the two states happens within the $\gamma=0$ photon sector meaning that there is not an actual absorption or emission of photons in the mixing process but rather both an absorption and subsequent emission (or vice versa) to and from a virtual $p$-like state in the $\gamma=1$ photon sector.
Due to the hydrogenic-like selection rules, there is no possible excitation/relaxation path between a bright $s$-like state and dark $p$-like, i.e. "d"-like states cannot function as virtual states, one do not get a finite oscillator strength. 
As demonstrated by the insets in Fig.~\ref{fig:fig3}(a) both the lower and upper branch of the DE polariton have a similar composition. 

While Fig.~\ref{fig:fig3}(a) illustrates how the cavity coupling reveals hidden (dark) states of a material, we observe another peculiarity in the polariton dispersion in Fig.~\ref{fig:fig3}(b), which corresponds to the circle (II) in Fig.~\ref{fig:fig2}(c). Through the cavity we can induce a quench of the optical activity within the polariton branch connecting two BEs, i.e. the spectral response vanishes completely between two consecutive BE lines. In this case, such a dimming of the polariton branch is \textit{not} originating from the contribution of DE states, but instead is the result of an interesting destructive interference between the optical transition amplitudes from the exciton lines connected by the photon branch. The destructive interference is demonstrated in the inset of Fig.~\ref{fig:fig3}(b) which reports the real and imaginary part of the transition amplitude from the groundstate to the $1s$ A and $1s$ B excitonic states, the states that make up most of the polariton state.

The QED-BSE method illustrated above requires considerable computational effort and it is therefore limited to few layers of lattice matched (commensurate) TMDs. The polaritonic physics can be well reproduced by replacing the BSE part in the QED-BSE with the Mott-Wannier (MW) equation, hence considerably reducing the computational cost and more importantly giving access to exciton-polariton physics in more complex materials such as encapsulated 2D crystals and van-der-Waals heterostructures. In this approach the bound excitons are described by an hydrogen like Schr\"odinger equation that provides the exciton envelope functions and hence allows the evaluation of the exciton-dipole matrix elements $\mathcal{M}^{\rm exc}_{mn}$ of the QED-BSE approach as explained in the supporting information. 

\begin{figure}
\centering
\includegraphics[width=1.0\columnwidth]{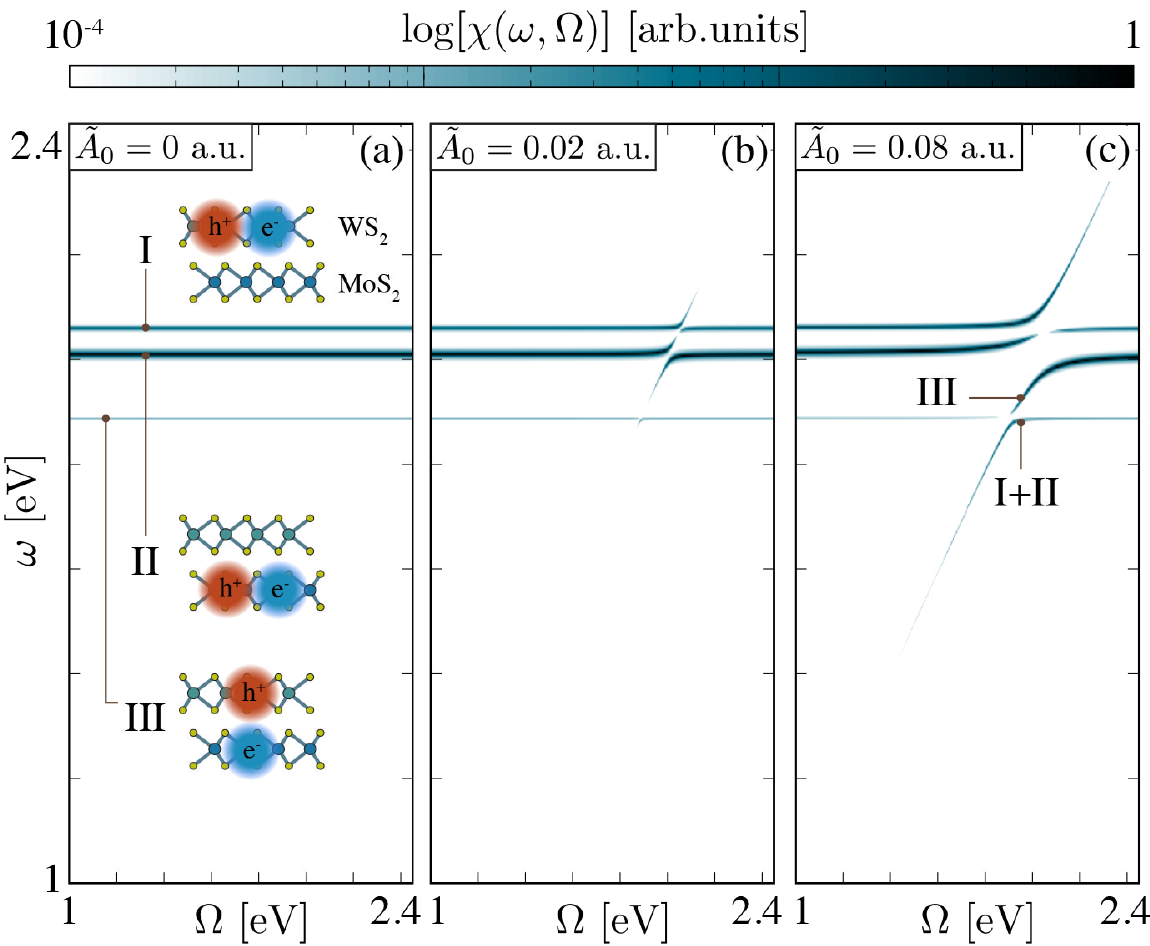}
   \caption{\label{fig:fig4}
   Exciton-polariton spectra of the MoS$_2$-WS$_2$ bilayer type-II vdWH as a function of cavity mode frequency $\Omega$ and for three different coupling strength $\tilde{A}_0$ in a.u.. Only the lowest bound inter- and intra-layer excitons are considered. Panel (a): Without coupling to the cavity mode the optical spectrum consists of non-dispersive optical spectral lines with interlayer exciton lying at lower transition energy than the intralayer exciton ones. Panel (b): For the weaker coupling both intra and inter-layer exciton lines undergo a Rabi splitting, however, due to the lower coupling with the cavity mode, the splitting for the interlayer excitons is much smaller (barely noticeable in the figure) than for the intra-layer ones. Panel (c): For the stronger coupling value the Rabi splitting for the interlayer excitons becomes sizable and the overall spectrum is drastically different from the uncoupled one. We stress that through the cavity mode we are able to invert the order of the intra- and inter-layer spectral lines. The broadening parameter used for the plots is $\eta=1.4$ meV.} 
\end{figure}

We apply the MW-QED to calculate the optical response of a vdWH consisting of MoS$_2$ and WS$_2$ bilayer. MoS$_2$ and WS$_2$ form a type-II heterostructure where the band edges of the two materials are aligned in a staggered fashion and which can host interlayer excitons along with intralayer ones in each of the two layers \cite{LatiniInterlayer2017, PalummoExciton2015}. For this particular bilayer choice, because of quasi-commensurability of the lattice of MoS$_2$ and WS$_2$, the QED-BSE has been used to further confirm the validity of the QED-MW presented in the following. However, for more general non-commensurate vdWHs the only feasible approach for simulating exciton-polariton properties is the QED-MW one.
The cavity frequency dependent optical response of such a structure for the most highly bound 1s intra- and inter-layer excitons is reported in Fig.~\ref{fig:fig4}. Including higher lying excitons does not affect the results presented here since the corresponding spectral features would appear in a higher energy window. The use of norm-conserving PAW-setups are needed to obtain the correct energy level alignment. Details on the method used to calculate the uncoupled excitonic resonances are discussed in the supplementary. 
Fig.~\ref{fig:fig4}(a) shows the optical spectrum of the vdWH without coupling with the cavity mode. Here, as indicated by the cartoons, the lower energy non-dispersive spectral line represents the optical transitions to the interlayer exciton whereas the higher energy ones are the transition to the intralayer excitons. The oscillator strength of the former is two order of magnitude lower than the one of the intralayer excitons, a direct consequence of the spatial separation of the bound electron-hole pair, justifying the weaker brightness observed in the spectrum. In Fig.~\ref{fig:fig4} the coupling with the cavity mode is switched on and the typical polaritonic branches appear with a sizable Rabi splitting for the intralayer and barely noticeable one for the interlayer one. The difference in Rabi splitting is, once again, a consequence of the weak interlayer exciton-light coupling. Similarly to the case of monolayer MoS$_2$ shown in Fig.~\ref{fig:fig3}(b) the interference between the transition to the two intralayer excitons results in a vanishing polariton branch. In the calculations, such an interference arises from the fact that the exciton wavefunctions of the two intralayer excitons appear with the same phase. While it is unsure whether this can be directly realized in experiment because any possible mechanism of dephasing would destroy the effect, one could potentially realize this by using external lasers to fix the relative phase of the different intralayer excitons.  
Finally, in the stronger coupling regime, shown in Fig.~\ref{fig:fig4}(c), the optical response is drastically modified and even the Rabi splitting of the interlayer exciton line is now appreciable. More interestingly, a similar analysis of the polaritonic wavefunctions as the one discussed for Fig.~\ref{fig:fig3}(c) shows that the energetic ordering of the transitions to intra and interlayer excitons is modified, as indicated in the panel. In particular the lower polariton branch at transition energies lower than the interlayer exciton resonance has now a non-negligible intralayer exciton composition. We stress that the reordering of the excitonic resonances is possible because of the different light-exciton coupling strength for the different types of excitons. From a practical point of view, the relative light-matter coupling strength for inter and intra-layer excitons can be controlled by (i) bulding active vdW-bilayers out of monolayers with different thicknesses and dielectric properties and (ii) choosing a different embedding material to change the dielectric environment of the cavity. For example a higher coupling strength for the interlayer excitons can be achieved in a vdW-bilayer made of crystals with a lower thickness and lower dielectric screening, whereas a lower coupling strength can be obtained choosing an embedding material with a higher dielectric function.  This calculations show the potential of cavity engineering for the optical design of functional devices where either intra-layer or inter-layer excitons can be excited on demand.

\begin{figure}
\centering
\includegraphics[width=1\columnwidth]{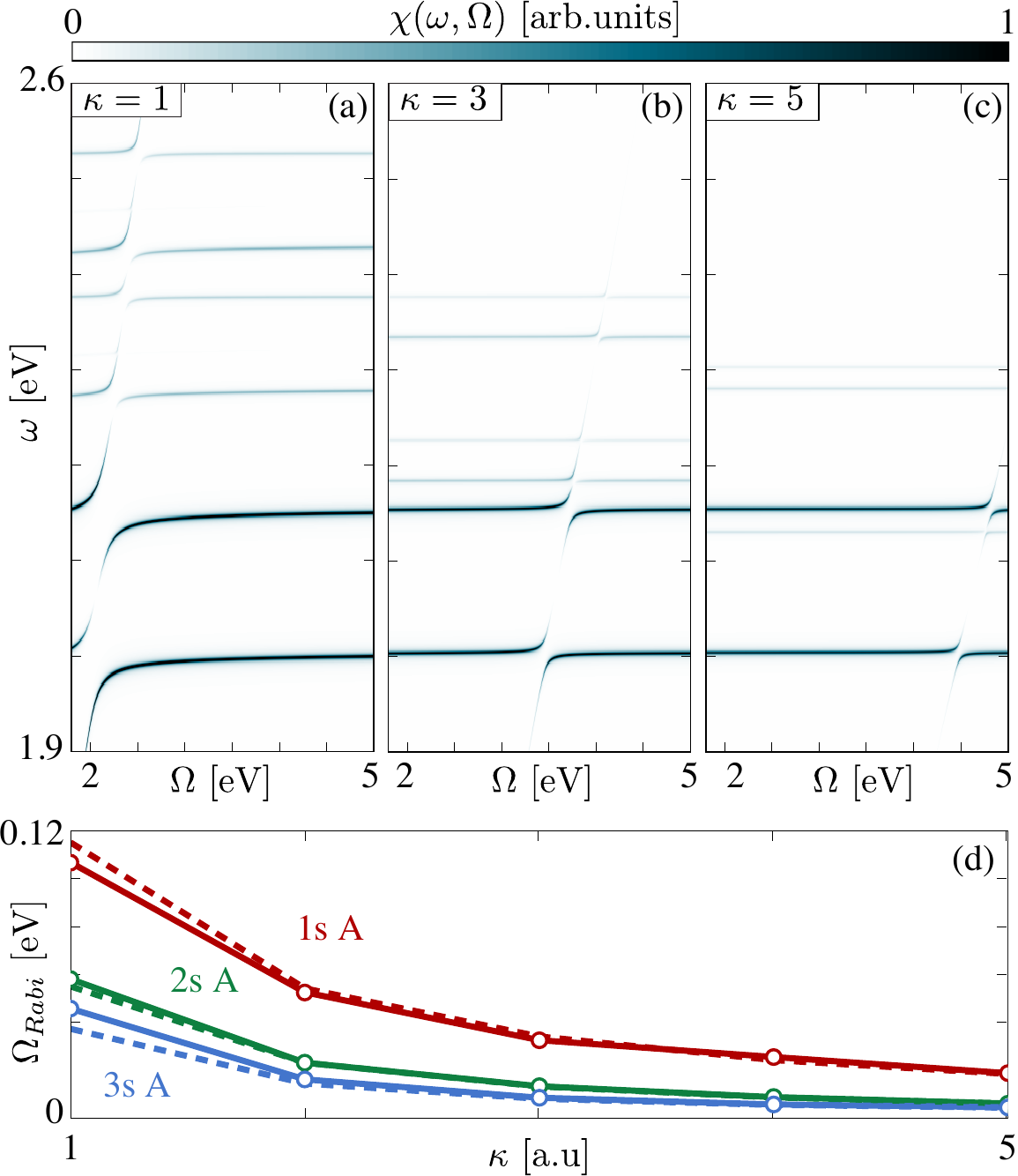}
   \caption{\label{fig:fig5}
 Polariton dispersion of MoS$_2$ for different values of dielectric constant $\kappa$ of the substrate (a-c), calculated with the MW model and with $\tilde{A}_0 =0.05$ a.u..
 Dependence of the Rabi splitting of the $1s$ A, $2s$ A and $3s$ A states on the dielectric constant (d). Solid lines with circles are the Rabi splitting calculated with the MW model, dashed lines are  calculated from the Jaynes-Cummings model as $\Omega_{Rabi} = 2A_0  \mathcal{M}^{\rm exc}_{0n}$.
    } 
\end{figure}

The MW-QED also provides a computationally inexpensive way of assessing the effect of embedding an active 2D materials in a bulk dielectric \cite{LatiniPhysRevB2015} (as sketched in Fig.~\ref{fig:fig1}). The effect of the dielectric environment is accounted for in the MW description through its bulk dielectric constant $\kappa$, c.f. refs.~\citenum{LatiniPhysRevB2015,andersen2015dielectric} and supplementary material for a detailed description of the method. The effect of environmental screening is threefold: (i) extra-screening of the electron-hole Coulomb interaction, (ii) renormalized photon dispersion ($\omega = \omega_c/\sqrt(\kappa)$) which now propagates in dielectric medium and (iii) by renormalizing the effective vector potential amplitude of the cavity $\bar{A}_0(\kappa)=\tilde{A}_0/\kappa$ \cite{GrossoBook2000}. 

In Fig.~\ref{fig:fig5}(a) we show the polariton dispersion of MoS$_2$ for fixed coupling strength $\tilde{A}_0$ and for various values of dielectric constants for the substrate material $\kappa$. The most striking feature is the change in the slope of the photon line due to the propagation in the cavity dielectric. This shifts the position of the polariton resonances. Furthermore, due to the change of the screened Coulomb interaction, the higher lying excitons are squeezed in energy towards the first excitonic state (cf. supplementary material).  
Fig.~\ref{fig:fig5}(c) shows the dependence of the Rabi-splitting of the first three BEs ($1s, 2s, 3s$ of the A series) as a function of the dielectric constant of the encapsulating material. We compare the splitting of the calculated dispersion with the Rabi splitting of the corresponding Jaynes-Cummings model where only two levels at a time (the groundstate and one of the excitons) are taken into account. Notably, only for small dielectric constants there are deviations from Jaynes-Cummings splitting indicating that for larger effective couplings $\bar{A}_0$ the two level approximation breaks down.

\paragraph*{Conclusion}
The QED-BSE formalism we presented here is a first principles approach to the properties of polaritons in optical cavities. We demostrate how it gives rise to complex polaritons dispersions, where hybridizations of polaritons manifest in features originating from DE. We also show how polaritonic interference could result in transparency and could be possibly achieved through laser control. We illustrate how the Mott-Wannier model for bound excitons can be used to obtain the relevant dipole matrix elements that enter the QED equations and thus provide a readily applicable method to compute such polariton properties. This method can be extended to account for the dielectric screening of materials inside the cavity which gives access to a large variety of van-der-Waals layered systems. In particular in the case of a type II vdWH, we demonstrate the possibility of engineering cavity-matter coupling and change the relative energy ordering of intra- and inter-layer excitons.
The QED-BSE and MW model presented here are the first ab-initio methods aimed at the design, understanding and control of the optoelectronic properties of materials embedded in quantum cavities.   

\section{Acknowledgements}
We are grateful for helpful discussions with Ch. Sch\"afer, M. Sentef and M. Ruggenthaler. S. L. acknowledges support from the Alexander von Humboldt foundation. We further acknowledge financial support from the European Research Council(ERC-2015-AdG-694097). The Flatiron Institute is a division of the Simons Foundation.

\begin{suppinfo}
Supporting Information Available: It contains extensive computational details, convergence tests with respect to the number of excitonic states, number of photons and number of cavity modes included in the calculation, details about the Mott-Wannier methodology and a comparison between the matter and photon spectral functions. This material is available free of charge via the Internet at http://pubs.acs.org.
\end{suppinfo}

\bibliography{bib}

\providecommand{\latin}[1]{#1}
\makeatletter
\providecommand{\doi}
  {\begingroup\let\do\@makeother\dospecials
  \catcode`\{=1 \catcode`\}=2 \doi@aux}
\providecommand{\doi@aux}[1]{\endgroup\texttt{#1}}
\makeatother
\providecommand*\mcitethebibliography{\thebibliography}
\csname @ifundefined\endcsname{endmcitethebibliography}
  {\let\endmcitethebibliography\endthebibliography}{}
\begin{mcitethebibliography}{50}
\providecommand*\natexlab[1]{#1}
\providecommand*\mciteSetBstSublistMode[1]{}
\providecommand*\mciteSetBstMaxWidthForm[2]{}
\providecommand*\mciteBstWouldAddEndPuncttrue
  {\def\EndOfBibitem{\unskip.}}
\providecommand*\mciteBstWouldAddEndPunctfalse
  {\let\EndOfBibitem\relax}
\providecommand*\mciteSetBstMidEndSepPunct[3]{}
\providecommand*\mciteSetBstSublistLabelBeginEnd[3]{}
\providecommand*\EndOfBibitem{}
\mciteSetBstSublistMode{f}
\mciteSetBstMaxWidthForm{subitem}{(\alph{mcitesubitemcount})}
\mciteSetBstSublistLabelBeginEnd
  {\mcitemaxwidthsubitemform\space}
  {\relax}
  {\relax}

\bibitem[Manzeli \latin{et~al.}(2017)Manzeli, Ovchinnikov, Pasquier, Yazyev,
  and Kis]{ManzeliNatureRevMat2017}
Manzeli,~S.; Ovchinnikov,~D.; Pasquier,~D.; Yazyev,~O.~V.; Kis,~A. \emph{Nat.
  Rev. Mater.} \textbf{2017}, \emph{2}, 17033\relax
\mciteBstWouldAddEndPuncttrue
\mciteSetBstMidEndSepPunct{\mcitedefaultmidpunct}
{\mcitedefaultendpunct}{\mcitedefaultseppunct}\relax
\EndOfBibitem
\bibitem[Wang \latin{et~al.}(2018)Wang, Chernikov, Glazov, Heinz, Marie, Amand,
  and Urbaszek]{WangRevModPhys2018}
Wang,~G.; Chernikov,~A.; Glazov,~M.~M.; Heinz,~T.~F.; Marie,~X.; Amand,~T.;
  Urbaszek,~B. \emph{Rev. Mod. Phys.} \textbf{2018}, \emph{90}, 021001\relax
\mciteBstWouldAddEndPuncttrue
\mciteSetBstMidEndSepPunct{\mcitedefaultmidpunct}
{\mcitedefaultendpunct}{\mcitedefaultseppunct}\relax
\EndOfBibitem
\bibitem[Ugeda \latin{et~al.}(2014)Ugeda, Bradley, Shi, da~Jornada, Zhang, Qiu,
  Ruan, Mo, Hussain, Shen, Wang, Louie, and Crommie]{UgedaNatureMater2014}
Ugeda,~M.~M.; Bradley,~A.~J.; Shi,~S.-F.; da~Jornada,~F.~H.; Zhang,~Y.;
  Qiu,~D.~Y.; Ruan,~W.; Mo,~S.-K.; Hussain,~Z.; Shen,~Z.-X. \latin{et~al.}
  \emph{Nat. Mater.} \textbf{2014}, \emph{13}, 1091\relax
\mciteBstWouldAddEndPuncttrue
\mciteSetBstMidEndSepPunct{\mcitedefaultmidpunct}
{\mcitedefaultendpunct}{\mcitedefaultseppunct}\relax
\EndOfBibitem
\bibitem[Ramasubramaniam(2012)]{RamasubramaniamPhysRevB2012}
Ramasubramaniam,~A. \emph{Phys. Rev. B} \textbf{2012}, \emph{86}, 115409\relax
\mciteBstWouldAddEndPuncttrue
\mciteSetBstMidEndSepPunct{\mcitedefaultmidpunct}
{\mcitedefaultendpunct}{\mcitedefaultseppunct}\relax
\EndOfBibitem
\bibitem[Cudazzo \latin{et~al.}(2011)Cudazzo, Tokatly, and
  Rubio]{CudazzoPhysRevB2011}
Cudazzo,~P.; Tokatly,~I.~V.; Rubio,~A. \emph{Phys. Rev. B} \textbf{2011},
  \emph{84}, 085406\relax
\mciteBstWouldAddEndPuncttrue
\mciteSetBstMidEndSepPunct{\mcitedefaultmidpunct}
{\mcitedefaultendpunct}{\mcitedefaultseppunct}\relax
\EndOfBibitem
\bibitem[Schaibley \latin{et~al.}(2016)Schaibley, Yu, Clark, Rivera, Ross,
  Seyler, Yao, and Xu]{SchaibleyNatureRevMat2016}
Schaibley,~J.~R.; Yu,~H.; Clark,~G.; Rivera,~P.; Ross,~J.~S.; Seyler,~K.~L.;
  Yao,~W.; Xu,~X. \emph{Nat. Rev. Mater.} \textbf{2016}, \emph{1}, 16055\relax
\mciteBstWouldAddEndPuncttrue
\mciteSetBstMidEndSepPunct{\mcitedefaultmidpunct}
{\mcitedefaultendpunct}{\mcitedefaultseppunct}\relax
\EndOfBibitem
\bibitem[Kim \latin{et~al.}(2017)Kim, Jin, Chen, Cai, Zhao, Lee, Kahn,
  Watanabe, Taniguchi, Tongay, Crommie, and Wang]{KimScienceAdv2017}
Kim,~J.; Jin,~C.; Chen,~B.; Cai,~H.; Zhao,~T.; Lee,~P.; Kahn,~S.; Watanabe,~K.;
  Taniguchi,~T.; Tongay,~S. \latin{et~al.}  \emph{Science Adv.} \textbf{2017},
  \emph{3}\relax
\mciteBstWouldAddEndPuncttrue
\mciteSetBstMidEndSepPunct{\mcitedefaultmidpunct}
{\mcitedefaultendpunct}{\mcitedefaultseppunct}\relax
\EndOfBibitem
\bibitem[Gao \latin{et~al.}(2012)Gao, Gao, Cannuccia, Taha-Tijerina, Balicas,
  Mathkar, Narayanan, Liu, Gupta, Peng, Yin, Rubio, and
  Ajayan]{GaoNanoLett2012}
Gao,~G.; Gao,~W.; Cannuccia,~E.; Taha-Tijerina,~J.; Balicas,~L.; Mathkar,~A.;
  Narayanan,~T.~N.; Liu,~Z.; Gupta,~B.~K.; Peng,~J. \latin{et~al.}  \emph{Nano
  Lett.} \textbf{2012}, \emph{12}, 3518--3525\relax
\mciteBstWouldAddEndPuncttrue
\mciteSetBstMidEndSepPunct{\mcitedefaultmidpunct}
{\mcitedefaultendpunct}{\mcitedefaultseppunct}\relax
\EndOfBibitem
\bibitem[Geim and Grigorieva(2013)Geim, and Grigorieva]{GeimNature2013}
Geim,~A.~K.; Grigorieva,~I.~V. \emph{Nature} \textbf{2013}, \emph{499},
  419\relax
\mciteBstWouldAddEndPuncttrue
\mciteSetBstMidEndSepPunct{\mcitedefaultmidpunct}
{\mcitedefaultendpunct}{\mcitedefaultseppunct}\relax
\EndOfBibitem
\bibitem[Hong \latin{et~al.}(2014)Hong, Kim, Shi, Zhang, Jin, Sun, Tongay, Wu,
  Zhang, and Wang]{HongUltrafast2014}
Hong,~X.; Kim,~J.; Shi,~S.-F.; Zhang,~Y.; Jin,~C.; Sun,~Y.; Tongay,~S.; Wu,~J.;
  Zhang,~Y.; Wang,~F. \emph{Nature nanotechnology} \textbf{2014}, \emph{9},
  682\relax
\mciteBstWouldAddEndPuncttrue
\mciteSetBstMidEndSepPunct{\mcitedefaultmidpunct}
{\mcitedefaultendpunct}{\mcitedefaultseppunct}\relax
\EndOfBibitem
\bibitem[Jin \latin{et~al.}(2018)Jin, Ma, Karni, Regan, Wang, and
  Heinz]{JinUltrafast2018}
Jin,~C.; Ma,~E.~Y.; Karni,~O.; Regan,~E.~C.; Wang,~F.; Heinz,~T.~F.
  \emph{Nature nanotechnology} \textbf{2018}, \emph{13}, 994\relax
\mciteBstWouldAddEndPuncttrue
\mciteSetBstMidEndSepPunct{\mcitedefaultmidpunct}
{\mcitedefaultendpunct}{\mcitedefaultseppunct}\relax
\EndOfBibitem
\bibitem[Flatten \latin{et~al.}(2016)Flatten, He, Coles, Trichet, Powell,
  Taylor, Warner, and Smith]{FlattenSciRep2016}
Flatten,~L.~C.; He,~Z.; Coles,~D.~M.; Trichet,~A. A.~P.; Powell,~A.~W.;
  Taylor,~R.~A.; Warner,~J.~H.; Smith,~J.~M. \emph{Sci. Rep.} \textbf{2016},
  \emph{6}, 33134\relax
\mciteBstWouldAddEndPuncttrue
\mciteSetBstMidEndSepPunct{\mcitedefaultmidpunct}
{\mcitedefaultendpunct}{\mcitedefaultseppunct}\relax
\EndOfBibitem
\bibitem[Chervy \latin{et~al.}(2018)Chervy, Azzini, Lorchat, Wang, Gorodetski,
  Hutchison, Berciaud, Ebbesen, and Genet]{ChervyACSPhot2018}
Chervy,~T.; Azzini,~S.; Lorchat,~E.; Wang,~S.; Gorodetski,~Y.;
  Hutchison,~J.~A.; Berciaud,~S.; Ebbesen,~T.~W.; Genet,~C. \emph{ACS
  Photonics} \textbf{2018}, \emph{5}, 1281--1287\relax
\mciteBstWouldAddEndPuncttrue
\mciteSetBstMidEndSepPunct{\mcitedefaultmidpunct}
{\mcitedefaultendpunct}{\mcitedefaultseppunct}\relax
\EndOfBibitem
\bibitem[Slootsky \latin{et~al.}(2014)Slootsky, Liu, Menon, and
  Forrest]{SlootskyPhysRevLett2014}
Slootsky,~M.; Liu,~X.; Menon,~V.~M.; Forrest,~S.~R. \emph{Phys. Rev. Lett.}
  \textbf{2014}, \emph{112}, 076401\relax
\mciteBstWouldAddEndPuncttrue
\mciteSetBstMidEndSepPunct{\mcitedefaultmidpunct}
{\mcitedefaultendpunct}{\mcitedefaultseppunct}\relax
\EndOfBibitem
\bibitem[Liu \latin{et~al.}(2014)Liu, Galfsky, Sun, Xia, Lin, Lee,
  K{\'e}na-Cohen, and Menon]{LiuNaturePhoton2014}
Liu,~X.; Galfsky,~T.; Sun,~Z.; Xia,~F.; Lin,~E.-c.; Lee,~Y.-H.;
  K{\'e}na-Cohen,~S.; Menon,~V.~M. \emph{Nat. Photonics} \textbf{2014},
  \emph{9}, 30\relax
\mciteBstWouldAddEndPuncttrue
\mciteSetBstMidEndSepPunct{\mcitedefaultmidpunct}
{\mcitedefaultendpunct}{\mcitedefaultseppunct}\relax
\EndOfBibitem
\bibitem[Sun \latin{et~al.}(2017)Sun, Gu, Ghazaryan, Shotan, Considine, Dollar,
  Chakraborty, Liu, Ghaemi, K{\'e}na-Cohen, and Menon]{SunNaturePhoton2017}
Sun,~Z.; Gu,~J.; Ghazaryan,~A.; Shotan,~Z.; Considine,~C.~R.; Dollar,~M.;
  Chakraborty,~B.; Liu,~X.; Ghaemi,~P.; K{\'e}na-Cohen,~S. \latin{et~al.}
  \emph{Nat. Photonics} \textbf{2017}, \emph{11}, 491\relax
\mciteBstWouldAddEndPuncttrue
\mciteSetBstMidEndSepPunct{\mcitedefaultmidpunct}
{\mcitedefaultendpunct}{\mcitedefaultseppunct}\relax
\EndOfBibitem
\bibitem[Flick \latin{et~al.}(2017)Flick, Ruggenthaler, Appel, and
  Rubio]{flick2017atoms}
Flick,~J.; Ruggenthaler,~M.; Appel,~H.; Rubio,~A. \emph{Proc. Nat. Ac. Sci.}
  \textbf{2017}, \relax
\mciteBstWouldAddEndPunctfalse
\mciteSetBstMidEndSepPunct{\mcitedefaultmidpunct}
{}{\mcitedefaultseppunct}\relax
\EndOfBibitem
\bibitem[Flick \latin{et~al.}(2018)Flick, Schäfer, Ruggenthaler, Appel, and
  Rubio]{flick2018ab}
Flick,~J.; Schäfer,~C.; Ruggenthaler,~M.; Appel,~H.; Rubio,~A. \emph{ACS
  Photonics} \textbf{2018}, \emph{5}, 992--1005\relax
\mciteBstWouldAddEndPuncttrue
\mciteSetBstMidEndSepPunct{\mcitedefaultmidpunct}
{\mcitedefaultendpunct}{\mcitedefaultseppunct}\relax
\EndOfBibitem
\bibitem[Ruggenthaler \latin{et~al.}(2018)Ruggenthaler, Tancogne-Dejean, Flick,
  Appel, and Rubio]{ruggenthaler2018quantum}
Ruggenthaler,~M.; Tancogne-Dejean,~N.; Flick,~J.; Appel,~H.; Rubio,~A.
  \emph{Nat. Rev. Chem.} \textbf{2018}, \emph{2}, 0118\relax
\mciteBstWouldAddEndPuncttrue
\mciteSetBstMidEndSepPunct{\mcitedefaultmidpunct}
{\mcitedefaultendpunct}{\mcitedefaultseppunct}\relax
\EndOfBibitem
\bibitem[Sheikhan \latin{et~al.}(2016)Sheikhan, Brennecke, and
  Kollath]{SheikhanPhysRevA2016}
Sheikhan,~A.; Brennecke,~F.; Kollath,~C. \emph{Phys. Rev. A} \textbf{2016},
  \emph{94}, 061603\relax
\mciteBstWouldAddEndPuncttrue
\mciteSetBstMidEndSepPunct{\mcitedefaultmidpunct}
{\mcitedefaultendpunct}{\mcitedefaultseppunct}\relax
\EndOfBibitem
\bibitem[Sentef \latin{et~al.}(2018)Sentef, Ruggenthaler, and
  Rubio]{Sentef:2018gp}
Sentef,~M.~A.; Ruggenthaler,~M.; Rubio,~A. \emph{Science Advances}
  \textbf{2018}, \emph{4}, eaau6969\relax
\mciteBstWouldAddEndPuncttrue
\mciteSetBstMidEndSepPunct{\mcitedefaultmidpunct}
{\mcitedefaultendpunct}{\mcitedefaultseppunct}\relax
\EndOfBibitem
\bibitem[Byrnes \latin{et~al.}(2014)Byrnes, Kim, and
  Yamamoto]{ByrnesNaturePhys2014}
Byrnes,~T.; Kim,~N.~Y.; Yamamoto,~Y. \emph{Nat. Phys.} \textbf{2014},
  \emph{10}, 803\relax
\mciteBstWouldAddEndPuncttrue
\mciteSetBstMidEndSepPunct{\mcitedefaultmidpunct}
{\mcitedefaultendpunct}{\mcitedefaultseppunct}\relax
\EndOfBibitem
\bibitem[Mazza and Georges(2019)Mazza, and Georges]{Mazza2019}
Mazza,~G.; Georges,~A. \emph{Physical Review Letters} \textbf{2019},
  \emph{122}, 017401\relax
\mciteBstWouldAddEndPuncttrue
\mciteSetBstMidEndSepPunct{\mcitedefaultmidpunct}
{\mcitedefaultendpunct}{\mcitedefaultseppunct}\relax
\EndOfBibitem
\bibitem[Salpeter and Bethe(1951)Salpeter, and Bethe]{SalpeterPhysRev1951}
Salpeter,~E.~E.; Bethe,~H.~A. \emph{Phys. Rev.} \textbf{1951}, \emph{84},
  1232--1242\relax
\mciteBstWouldAddEndPuncttrue
\mciteSetBstMidEndSepPunct{\mcitedefaultmidpunct}
{\mcitedefaultendpunct}{\mcitedefaultseppunct}\relax
\EndOfBibitem
\bibitem[Rohlfing and Louie(1998)Rohlfing, and Louie]{RohlfingPhysRevLett1998}
Rohlfing,~M.; Louie,~S.~G. \emph{Phys. Rev. Lett.} \textbf{1998}, \emph{81},
  2312--2315\relax
\mciteBstWouldAddEndPuncttrue
\mciteSetBstMidEndSepPunct{\mcitedefaultmidpunct}
{\mcitedefaultendpunct}{\mcitedefaultseppunct}\relax
\EndOfBibitem
\bibitem[Rohlfing and Louie(2000)Rohlfing, and Louie]{RohlfingPhysRevB2000}
Rohlfing,~M.; Louie,~S.~G. \emph{Phys. Rev. B} \textbf{2000}, \emph{62},
  4927--4944\relax
\mciteBstWouldAddEndPuncttrue
\mciteSetBstMidEndSepPunct{\mcitedefaultmidpunct}
{\mcitedefaultendpunct}{\mcitedefaultseppunct}\relax
\EndOfBibitem
\bibitem[Onida \latin{et~al.}(2002)Onida, Reining, and
  Rubio]{OnidaRevModPhys2002}
Onida,~G.; Reining,~L.; Rubio,~A. \emph{Rev. Mod. Phys.} \textbf{2002},
  \emph{74}, 601--659\relax
\mciteBstWouldAddEndPuncttrue
\mciteSetBstMidEndSepPunct{\mcitedefaultmidpunct}
{\mcitedefaultendpunct}{\mcitedefaultseppunct}\relax
\EndOfBibitem
\bibitem[Tokatly(2013)]{Tokatly:2013co}
Tokatly,~I.~V. \emph{Phys. Rev. Lett.} \textbf{2013}, \emph{110}, 233001\relax
\mciteBstWouldAddEndPuncttrue
\mciteSetBstMidEndSepPunct{\mcitedefaultmidpunct}
{\mcitedefaultendpunct}{\mcitedefaultseppunct}\relax
\EndOfBibitem
\bibitem[Ruggenthaler \latin{et~al.}(2014)Ruggenthaler, Flick, Pellegrini,
  Appel, Tokatly, and Rubio]{RuggenthalerPhysRevA2014}
Ruggenthaler,~M.; Flick,~J.; Pellegrini,~C.; Appel,~H.; Tokatly,~I.~V.;
  Rubio,~A. \emph{Phys. Rev. A} \textbf{2014}, \emph{90}, 012508\relax
\mciteBstWouldAddEndPuncttrue
\mciteSetBstMidEndSepPunct{\mcitedefaultmidpunct}
{\mcitedefaultendpunct}{\mcitedefaultseppunct}\relax
\EndOfBibitem
\bibitem[Pellegrini \latin{et~al.}(2015)Pellegrini, Flick, Tokatly, Appel, and
  Rubio]{Pellegrini:2015kq}
Pellegrini,~C.; Flick,~J.; Tokatly,~I.~V.; Appel,~H.; Rubio,~A. \emph{Phys.
  Rev. Lett.} \textbf{2015}, \emph{115}, 093001\relax
\mciteBstWouldAddEndPuncttrue
\mciteSetBstMidEndSepPunct{\mcitedefaultmidpunct}
{\mcitedefaultendpunct}{\mcitedefaultseppunct}\relax
\EndOfBibitem
\bibitem[Flick \latin{et~al.}(2015)Flick, Ruggenthaler, Appel, and
  Rubio]{FlickProcNatAcSci2015}
Flick,~J.; Ruggenthaler,~M.; Appel,~H.; Rubio,~A. \emph{Proc. Nat. Ac. Sci.}
  \textbf{2015}, \emph{112}, 15285--15290\relax
\mciteBstWouldAddEndPuncttrue
\mciteSetBstMidEndSepPunct{\mcitedefaultmidpunct}
{\mcitedefaultendpunct}{\mcitedefaultseppunct}\relax
\EndOfBibitem
\bibitem[Rokaj \latin{et~al.}(2018)Rokaj, Welakuh, Ruggenthaler, and
  Rubio]{Rokaj:2018dn}
Rokaj,~V.; Welakuh,~D.~M.; Ruggenthaler,~M.; Rubio,~A. \emph{J. Phys. B: At.
  Mol. Opt. Phys.} \textbf{2018}, \emph{51}, 034005\relax
\mciteBstWouldAddEndPuncttrue
\mciteSetBstMidEndSepPunct{\mcitedefaultmidpunct}
{\mcitedefaultendpunct}{\mcitedefaultseppunct}\relax
\EndOfBibitem
\bibitem[{Flick} \latin{et~al.}(2018){Flick}, {Welakuh}, {Ruggenthaler},
  {Appel}, and {Rubio}]{DavisArXiv2018}
{Flick},~J.; {Welakuh},~D.~M.; {Ruggenthaler},~M.; {Appel},~H.; {Rubio},~A.
  \emph{ArXiv e-prints} \textbf{2018}, arXiv:1803.02519\relax
\mciteBstWouldAddEndPuncttrue
\mciteSetBstMidEndSepPunct{\mcitedefaultmidpunct}
{\mcitedefaultendpunct}{\mcitedefaultseppunct}\relax
\EndOfBibitem
\bibitem[Du \latin{et~al.}(2018)Du, Martínez-Martínez, Ribeiro, Hu, Menon,
  and Yuen-Zhou]{DuChemSci2018}
Du,~M.; Martínez-Martínez,~L.~A.; Ribeiro,~R.~F.; Hu,~Z.; Menon,~V.~M.;
  Yuen-Zhou,~J. \emph{Chem. Sci.} \textbf{2018}, \emph{9}, 6659--6669\relax
\mciteBstWouldAddEndPuncttrue
\mciteSetBstMidEndSepPunct{\mcitedefaultmidpunct}
{\mcitedefaultendpunct}{\mcitedefaultseppunct}\relax
\EndOfBibitem
\bibitem[{Polak} \latin{et~al.}(2018){Polak}, {Jayaprakash}, {Leventis},
  {Fallon}, {Coulthard}, {Petty}, {Anthony}, {Bronstein}, {Lidzey}, {Clark},
  and {Musser}]{PolakArXiv2018}
{Polak},~D.; {Jayaprakash},~R.; {Leventis},~A.; {Fallon},~K.~J.;
  {Coulthard},~H.; {Petty},~I.,~Anthony~J.; {Anthony},~J.; {Bronstein},~H.;
  {Lidzey},~D.~G.; {Clark},~J. \latin{et~al.}  \emph{arXiv e-prints}
  \textbf{2018}, arXiv:1806.09990\relax
\mciteBstWouldAddEndPuncttrue
\mciteSetBstMidEndSepPunct{\mcitedefaultmidpunct}
{\mcitedefaultendpunct}{\mcitedefaultseppunct}\relax
\EndOfBibitem
\bibitem[Cao \latin{et~al.}(2018)Cao, Wu, and Louie]{CaoPhysRevLett2018}
Cao,~T.; Wu,~M.; Louie,~S.~G. \emph{Phys. Rev. Lett.} \textbf{2018},
  \emph{120}, 087402\relax
\mciteBstWouldAddEndPuncttrue
\mciteSetBstMidEndSepPunct{\mcitedefaultmidpunct}
{\mcitedefaultendpunct}{\mcitedefaultseppunct}\relax
\EndOfBibitem
\bibitem[Ye \latin{et~al.}(2014)Ye, Cao, O{\textquoteright}Brien, Zhu, Yin,
  Wang, Louie, and Zhang]{YeNature2014}
Ye,~Z.; Cao,~T.; O{\textquoteright}Brien,~K.; Zhu,~H.; Yin,~X.; Wang,~Y.;
  Louie,~S.~G.; Zhang,~X. \emph{Nature} \textbf{2014}, \emph{513},
  214--218\relax
\mciteBstWouldAddEndPuncttrue
\mciteSetBstMidEndSepPunct{\mcitedefaultmidpunct}
{\mcitedefaultendpunct}{\mcitedefaultseppunct}\relax
\EndOfBibitem
\bibitem[Loh(2017)]{Loh:2017cw}
Loh,~K.~P. \emph{Nature Nanotechnology} \textbf{2017}, \emph{12},
  837--838\relax
\mciteBstWouldAddEndPuncttrue
\mciteSetBstMidEndSepPunct{\mcitedefaultmidpunct}
{\mcitedefaultendpunct}{\mcitedefaultseppunct}\relax
\EndOfBibitem
\bibitem[Latini \latin{et~al.}(2017)Latini, Winther, Olsen, and
  Thygesen]{LatiniInterlayer2017}
Latini,~S.; Winther,~K.~T.; Olsen,~T.; Thygesen,~K.~S. \emph{Nano letters}
  \textbf{2017}, \emph{17}, 938--945\relax
\mciteBstWouldAddEndPuncttrue
\mciteSetBstMidEndSepPunct{\mcitedefaultmidpunct}
{\mcitedefaultendpunct}{\mcitedefaultseppunct}\relax
\EndOfBibitem
\bibitem[Grosso and Parravicini(2000)Grosso, and Parravicini]{GrossoBook2000}
Grosso,~G.; Parravicini,~G. \emph{Solid State Physics}; Elsevier Science,
  2000\relax
\mciteBstWouldAddEndPuncttrue
\mciteSetBstMidEndSepPunct{\mcitedefaultmidpunct}
{\mcitedefaultendpunct}{\mcitedefaultseppunct}\relax
\EndOfBibitem
\bibitem[Keldysh(1979)]{KeldyshJETPLett1979}
Keldysh,~L.~V. \emph{JETP Lett.} \textbf{1979}, \emph{29}, 658\relax
\mciteBstWouldAddEndPuncttrue
\mciteSetBstMidEndSepPunct{\mcitedefaultmidpunct}
{\mcitedefaultendpunct}{\mcitedefaultseppunct}\relax
\EndOfBibitem
\bibitem[Latini \latin{et~al.}(2015)Latini, Olsen, and
  Thygesen]{LatiniPhysRevB2015}
Latini,~S.; Olsen,~T.; Thygesen,~K.~S. \emph{Phys. Rev. B} \textbf{2015},
  \emph{92}, 245123\relax
\mciteBstWouldAddEndPuncttrue
\mciteSetBstMidEndSepPunct{\mcitedefaultmidpunct}
{\mcitedefaultendpunct}{\mcitedefaultseppunct}\relax
\EndOfBibitem
\bibitem[Mortensen \latin{et~al.}(2005)Mortensen, Hansen, and
  Jacobsen]{MortensenPhysRevB2005}
Mortensen,~J.~J.; Hansen,~L.~B.; Jacobsen,~K.~W. \emph{Phys. Rev. B}
  \textbf{2005}, \emph{71}, 035109\relax
\mciteBstWouldAddEndPuncttrue
\mciteSetBstMidEndSepPunct{\mcitedefaultmidpunct}
{\mcitedefaultendpunct}{\mcitedefaultseppunct}\relax
\EndOfBibitem
\bibitem[Enkovaara \latin{et~al.}(2010)Enkovaara, Rostgaard, Mortensen, Chen,
  Du{\l}ak, Ferrighi, Gavnholt, Glinsvad, Haikola, Hansen, Kristoffersen,
  Kuisma, Larsen, Lehtovaara, Ljungberg, Lopez-Acevedo, Moses, Ojanen, Olsen,
  Petzold, Romero, Stausholm-M{\o}ller, Strange, Tritsaris, Vanin, Walter,
  Hammer, Häkkinen, Madsen, Nieminen, N{\o}rskov, Puska, Rantala, Schi{\o}tz,
  Thygesen, and Jacobsen]{EnkovaaraJPhysCondMat2010}
Enkovaara,~J.; Rostgaard,~C.; Mortensen,~J.~J.; Chen,~J.; Du{\l}ak,~M.;
  Ferrighi,~L.; Gavnholt,~J.; Glinsvad,~C.; Haikola,~V.; Hansen,~H.~A.
  \latin{et~al.}  \emph{J. Phys.: Cond. Mat.} \textbf{2010}, \emph{22},
  253202\relax
\mciteBstWouldAddEndPuncttrue
\mciteSetBstMidEndSepPunct{\mcitedefaultmidpunct}
{\mcitedefaultendpunct}{\mcitedefaultseppunct}\relax
\EndOfBibitem
\bibitem[Flick \latin{et~al.}(2018)Flick, Rivera, and
  Narang]{FlickNanophotonics2018}
Flick,~J.; Rivera,~N.; Narang,~P. \emph{Nanophotonics} \textbf{2018}, \emph{7},
  1479--1501\relax
\mciteBstWouldAddEndPuncttrue
\mciteSetBstMidEndSepPunct{\mcitedefaultmidpunct}
{\mcitedefaultendpunct}{\mcitedefaultseppunct}\relax
\EndOfBibitem
\bibitem[Qiu \latin{et~al.}(2013)Qiu, Felipe, and Louie]{qiu2013optical}
Qiu,~D.~Y.; Felipe,~H.; Louie,~S.~G. \emph{Phys. Rev. Lett.} \textbf{2013},
  \emph{111}, 216805\relax
\mciteBstWouldAddEndPuncttrue
\mciteSetBstMidEndSepPunct{\mcitedefaultmidpunct}
{\mcitedefaultendpunct}{\mcitedefaultseppunct}\relax
\EndOfBibitem
\bibitem[Malic \latin{et~al.}(2018)Malic, Selig, Feierabend, Brem,
  Christiansen, Wendler, Knorr, and Bergh\"auser]{MalicPhysRevMat2018}
Malic,~E.; Selig,~M.; Feierabend,~M.; Brem,~S.; Christiansen,~D.; Wendler,~F.;
  Knorr,~A.; Bergh\"auser,~G. \emph{Phys. Rev. Materials} \textbf{2018},
  \emph{2}, 014002\relax
\mciteBstWouldAddEndPuncttrue
\mciteSetBstMidEndSepPunct{\mcitedefaultmidpunct}
{\mcitedefaultendpunct}{\mcitedefaultseppunct}\relax
\EndOfBibitem
\bibitem[Palummo \latin{et~al.}(2015)Palummo, Bernardi, and
  Grossman]{PalummoExciton2015}
Palummo,~M.; Bernardi,~M.; Grossman,~J.~C. \emph{Nano letters} \textbf{2015},
  \emph{15}, 2794--2800\relax
\mciteBstWouldAddEndPuncttrue
\mciteSetBstMidEndSepPunct{\mcitedefaultmidpunct}
{\mcitedefaultendpunct}{\mcitedefaultseppunct}\relax
\EndOfBibitem
\bibitem[Andersen \latin{et~al.}(2015)Andersen, Latini, and
  Thygesen]{andersen2015dielectric}
Andersen,~K.; Latini,~S.; Thygesen,~K.~S. \emph{Nano Lett.} \textbf{2015},
  \emph{15}, 4616--4621\relax
\mciteBstWouldAddEndPuncttrue
\mciteSetBstMidEndSepPunct{\mcitedefaultmidpunct}
{\mcitedefaultendpunct}{\mcitedefaultseppunct}\relax
\EndOfBibitem
\end{mcitethebibliography}
\clearpage



\section*{Supplementary material}

# Supporting Information: Cavity control of Excitons in two dimensional Materials

Simone Latini,[*,†,§] Enrico Ronca,[*,†,§] Umberto De Giovannini,[†,‡]

Hannes Hübener,[†] and Angel Rubio[*,†,¶]

[†]*Max Planck Institute for the Structure and Dynamics of Matter, Luruper Chaussee 149, 22761 Hamburg, Germany.*

*Center for Free-Electron Laser Science and Department of Physics, University of Hamburg, Luruper Chaussee 149, 22761 Hamburg, Germany*

[‡]*Dipartimento di Fisica e Chimica, Universitá degli Studi di Palermo, Via Archirafi 36, I-90123, Palermo, Italy*

[¶]*Center for Computational Quantum Physics (CCQ), The Flatiron Institute, 162 Fifth avenue, New York NY 10010.*

[§]*Contributed equally to this work*

E-mail: simone.latini@mpsd.mpg.de; enrico.ronca@mpsd.mpg.de; angel.rubio@mpsd.mpg.de

## QED-BSE Computational Details

All of the ab-initio calculations for the electronic problem in this work are performed with the GPAW code.[1,2] The single particle energies and wavefunctions, together with the momentum matrix elements $\langle \phi_{i\mathbf{k}} | \hat{e} \cdot \hat{p} | \phi_{j\mathbf{k}} \rangle$ are calculated within density functional theory with the LDA exchange correlation functional on a plane-wave basis. The LDA calculations for the monolayers TMDs were performed using a plane wave basis set with a cut off energy of 500 eV and $60 \times 60$ k-point grids.

We calculate the excitonic wavefunctions used in the QED Hamiltonian by solving the BSE considering only electron-hole pairs formed between the top valence and bottom conduction band. We take a cut-off energy of 150 eV for the evaluation of the screened interaction, which is calculated within the RPA on LDA energies and wavefunctions. A scissor operator based on the $G_0W_0$ calculations is applied for a better description of the electronic gap, see Tab. S1 for specific values. Spin-orbit effects are included post BSE solution by including two equivalent spin-orthogonal (A and B) excitonic series. Spin-orbit splitting values are reported in Tab. S1. The Brillouin zone sampling for BSE and $G_0W_0$ calculations is done on a $60 \times 60$ k-point-mesh.

Table S1: GW band gap and SOC splittings for different TMDs. All values are extracted from the C2DB two-dimensional materials database[3]

| TMD | GW gap (eV) | SOC split (eV) |
|---|---|---|
| $MoS_2$ | 2.53 | 0.15 |
| $MoSe_2$ | 2.14 | 0.17 |
| $WS_2$ | 2.60 | 0.37 |
| $WSe_2$ | 2.21 | 0.40 |

Finally, in the QED Hamiltonian 18 excitonic states (up to the $3s$), a single photon mode with photon number states up to $\gamma = 3$ are included. These values have been chosen as result of convergence tests on the matter polarizability spectra in the energy region and $\tilde{A}_0$ values investigated here. Convergence with respect to the number of modes is discussed in more details in the following section. Exciton-polariton dispersions for $MoS_2$, $WS_2$, $MoSe_2$ and $WSe_2$ are shown in Fig. S1

## QED-BSE Multi-mode Calculations

To avoid unnecessary complications, in the main text we only treated a single cavity mode. Eq. (1) of the main text can be directly generalized to the case of multiple modes, in formulas:

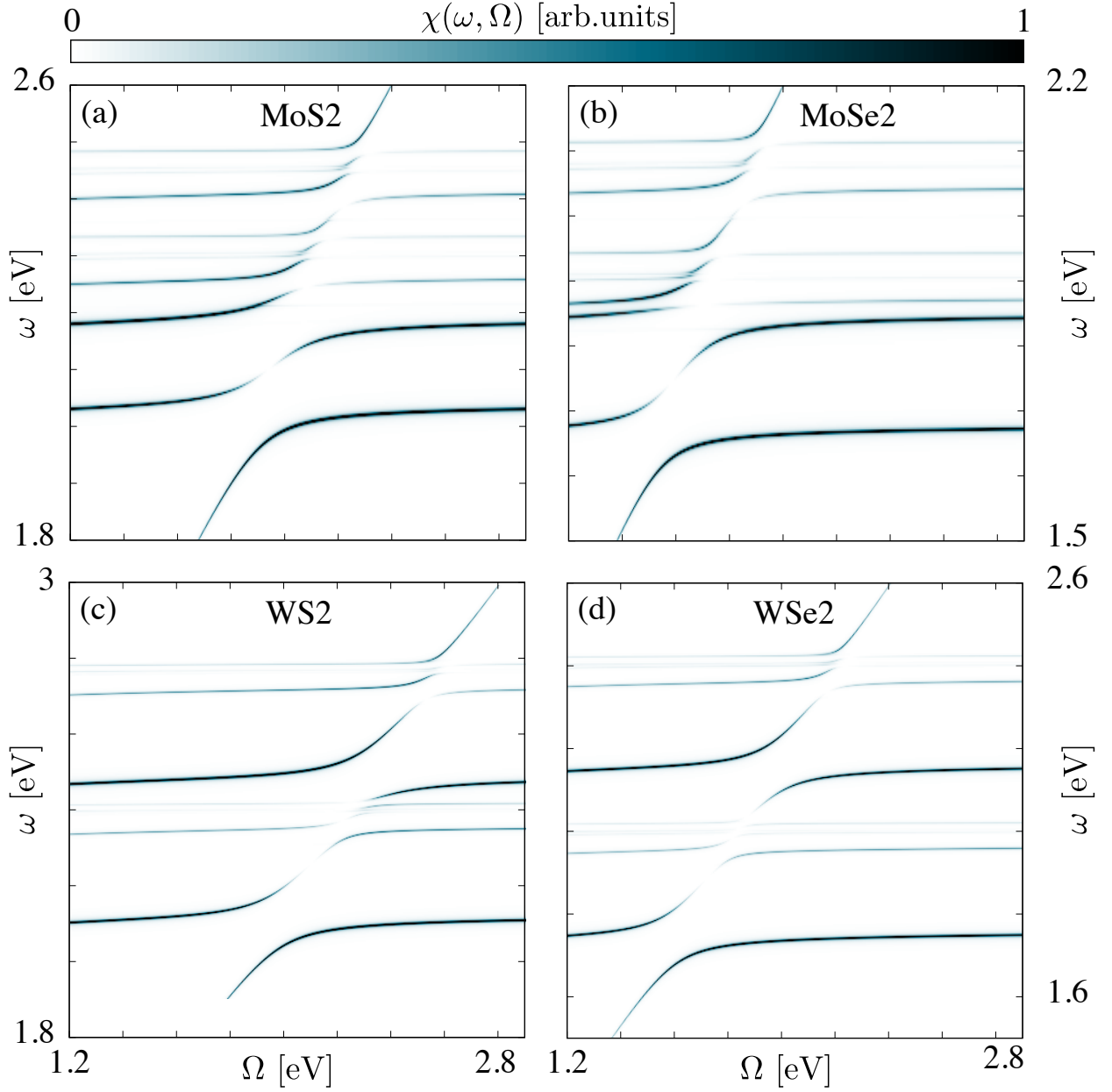

Figure S1: Exciton polariton dispersion of different TMDs with a coupling constant of $\tilde{A}_0$=0.08 a.u..

$$\begin{aligned}\hat{H}_{\text{QED}} =& \hat{H}_{\text{el}} + \sum_{\mathbf{q}\sigma} \Omega_q \hat{a}^\dagger_{\mathbf{q}\sigma} \hat{a}_{\mathbf{q}\sigma} + \\ & \frac{N_{\text{el}}}{2} \sum_{\mathbf{q}\sigma} A^2_{0\mathbf{q}\sigma} \left|(u^*_{\mathbf{q}\sigma}\hat{a}^\dagger_{\mathbf{q}\sigma} + u_{\mathbf{q}\sigma}\hat{a}_{\mathbf{q}\sigma})\right|^2 + \\ & \sum_{\mathbf{q}\sigma} A_{0\mathbf{q}\sigma} \sum_{ij\mathbf{k}\mathbf{k}'} \left( \langle \phi_{i\mathbf{k}} | u^*_{\mathbf{q}\sigma} \hat{p}_\sigma | \phi_{j\mathbf{k}'} \rangle \hat{d}^\dagger_{i\mathbf{k}} \hat{d}_{j\mathbf{k}'} \hat{a}^\dagger_{\mathbf{q}\sigma} + h.c. \right),\end{aligned} \quad (1)$$

3

where $\mathbf{q}$ is the photon momentum, $\sigma$ the $x, y, z$ index and $u_{\mathbf{q}\sigma}$ the photon polarization vectors. In a quasi-2D cavity the latter are given by:

$$\begin{aligned} u_{\mathbf{q}x} = u_{\mathbf{q}y} &= e^{i\mathbf{q}_\parallel \cdot \mathbf{r}_\parallel} i \sin(q_\perp z), \\ u_{\mathbf{q}z} &= e^{i\mathbf{q}_\parallel \cdot \mathbf{r}_\parallel} \cos(q_\perp z). \end{aligned} \qquad (2)$$

In our work the relevant quantization dimension is the out-of-plane one and we only consider photon modes for which $\mathbf{q}_\parallel = 0$. We then assume that the active 2D crystal has an infinitesimal thickness and it is placed at the center of the cavity, namely $z = L_\perp/2$. Given the isotropy of TMDs monolayers, we can choose the in plane polarization to be in the y-direction, without any loss of generality. In such a setting and because 2D excitons can only couple to in-plane electric fields we consider only polarization of the kind $u_{\mathbf{q}y} = i\sin(q_\perp L_\perp/2)$ and of these only the odd modes are non-zero, i.e. $q_\perp = \frac{\pi\alpha}{L_\perp}$ with $\alpha = 1, 3, 5, ....$ With the restriction of the modes just described, the long wavelenght approximation can be still applied and we can simplify the Hamiltonian in Eq. 1 as:

$$\begin{aligned} \hat{H}_{\text{QED}} =& \hat{H}_{\text{el}} + \sum_\alpha \Omega_\alpha \hat{a}^\dagger_\alpha \hat{a}_\alpha + \frac{N_{\text{el}}}{2} \sum_\alpha A_{0\alpha}^2 \left|(\hat{a}^\dagger_\alpha + \hat{a}_\alpha)\right|^2 + \\ & \sum_\alpha A_{0\alpha} \sum_{ij\mathbf{k}} \left( \langle \phi_{i\mathbf{k}} | \hat{e} \cdot \hat{p} | \phi_{j\mathbf{k}} \rangle \hat{d}^\dagger_{i\mathbf{k}} \hat{d}_{j\mathbf{k}} \hat{a}^\dagger_\alpha + h.c. \right), \end{aligned} \qquad (3)$$

with $\Omega_\alpha = \alpha\pi c/L_\perp$ and $A_{0\alpha} = 1/\sqrt{2\pi c S}\sqrt{\alpha}$.

Following the procedure we used for the single mode case, we project the Hamiltonian above onto the product state basis $|\Psi_n^{\text{exc}}\rangle \otimes |\gamma^{\alpha=1}\rangle \otimes |\gamma^{\alpha=3}\rangle \otimes \cdots$, where $|\gamma^\alpha\rangle$ are the eigenfunctions of the photonic harmonic oscillator associated with the different modes. The Hamiltonian can be then diagonalized and the resulting polaritonic states and energy can be used for calculating the optical response as in Eq. (5) of the main text. Calculations for MoS$_2$ with up to $\alpha = 9$ are reported in Fig. S2. Within the $\Omega$ range considered in the main text the multi-mode calculation is in quantitative agreement with the single-mode one. However including more modes allows us to explore smaller values of $\Omega$, where polaritonic

effects due to the higher modes take place.

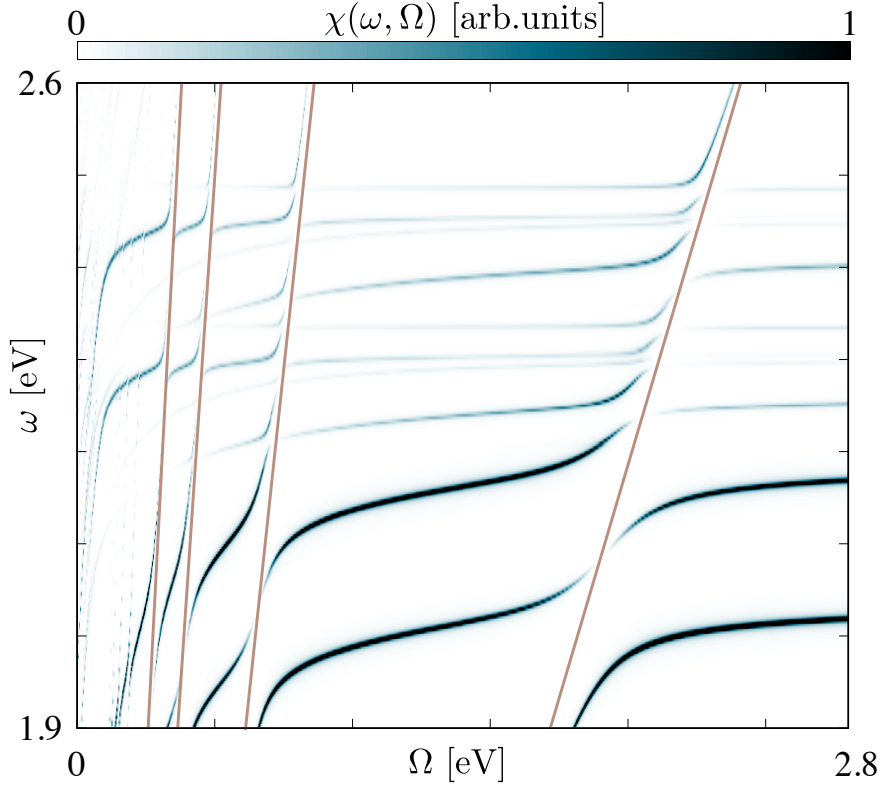

Figure S2: Exciton-polariton spectra of $MoS_2$ in a cavity as a function of cavity mode frequency $\Omega$ obtained with a multi-mode model with $\alpha = 9$ modes and a coupling strength of $\tilde{A}_0 = 0.08$ a.u..

**Photon response**

Equivalently to eq. 5 in the main text, a response function for photons can be also formulated:

$$\chi^{phot}(\omega, \Omega) = \sum_I \frac{(\tilde{\mathcal{M}}_{I0}^{\text{pol}})^* \tilde{\mathcal{M}}_{I0}^{\text{pol}}}{\omega - E_I^{\text{pol}}(\Omega) + E_0^{\text{pol}}(\Omega) + i\eta}. \qquad (4)$$

In this case, the matrix elements are defined as $\tilde{\mathcal{M}}_{IJ}^{\text{pol}} = \langle \Psi_I^{\text{pol}} | \hat{a}^\dagger | \Psi_J^{\text{pol}} \rangle = \sum_{n\gamma\theta} C_{n\gamma}^{I*} C_{m\theta}^{J} \mathcal{M}_{\gamma\theta}^{\text{phot}}$ where $\mathcal{M}_{\gamma\theta}^{\text{phot}} = \langle \gamma | \hat{a}^\dagger | \theta \rangle$. Such a quantity represents the photonic counterpart of the matter polarizability and it is of relevance for experiments where the photons are probed. The comparison between the matter and photon polarizabilities of $MoS_2$ for the stronger coupling values is presented in Fig. S3. The two response functions provide complementary informa-

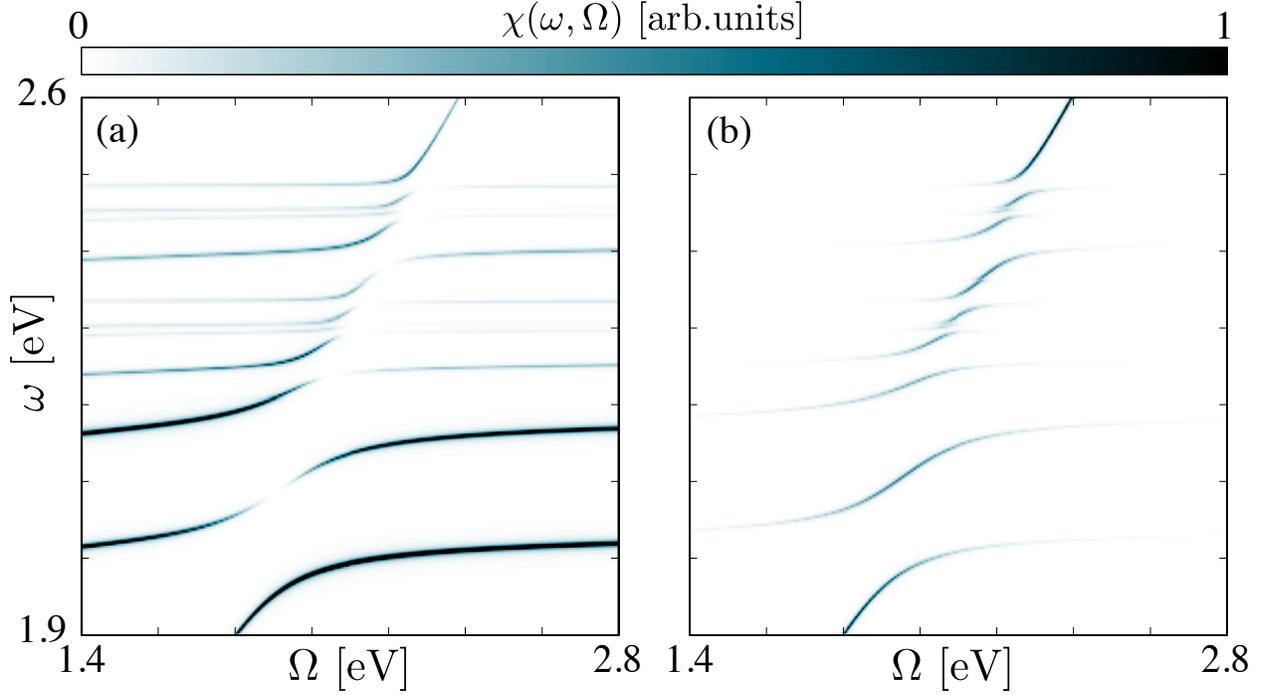

Figure S3: Comparison between (a) matter and (b) photon polarizabilitues for $MoS_2$ for a coupling value $\tilde{A}_0 = 0.08$ a.u..

tion on the polaritonic response, with the photonic response function showing a feature-rich matter-dressing of the bare photon line.

## MW Method and Computational Details

In the MW model, the exciton is treated as a hydrogen atom, where the electron and the hole interact via a screened Coulomb interaction, described by the Schrödinger equation[4]

$$\left[-\frac{\nabla^2}{2\mu} - W(\mathbf{r})\right] F^n(\mathbf{r}) = E_{b,\text{exc}}^n F^n(\mathbf{r}). \tag{5}$$

with $\mu$ the exciton effective mass and the screened Coulomb interaction

$$W(\mathbf{r}) = \frac{1}{4\alpha} \left[H_0(x) - N_0(x)\right]_{x=r/2\pi\alpha}, \tag{6}$$

where $H_0$ and $N_0$ are the Struve and Neumann functions and $\alpha$ is the crystal polarizability. By solving Eq. (5), we obtain exciton binding energies ($E_{b,\text{exc}}^n = E_{\text{gap}} - E_{\text{exc}}^n$) and real space excitonic functions $F^n(\mathbf{r})$. Within the MW model, the latter represent the Fourier transforms of the envelope functions $A_{\mathbf{k}}^n$, i.e.

$$F^n(\mathbf{r}) = \frac{1}{N_k} \sum_{\mathbf{k}} A_{\mathbf{k}}^n e^{i\mathbf{k}\cdot\mathbf{r}}. \qquad (7)$$

We stress that the coefficients $A_{\mathbf{k}}^n$ used in this section are the MW approximation to those defined above for BSE. An extensive discussion on the MW approach, the related computational advantages and how it can be derived as an approximation of the BSE can be found in Ref.[4]

Borrowing the assumptions of the MW model, in particular that the excitons are extremely localized at certain points in the Brillouin zone, the $K$-points for TMDs, we assume that the valence-conduction momentum matrix element is constant, and simplify Eq. (3) of the main text to:

$$\mathcal{M}_{0n}^{\text{exc}} = \langle \phi_{v\text{K}} | \hat{e} \cdot \hat{p} | \phi_{c\text{K}} \rangle F^n(\mathbf{r}=0). \qquad (8)$$

This shows that whether an exciton is bright or dark, is determined by the value of its real space wavefunction at the origin. This explains why only $s$-like excitons are bright. Simplifications can also be done for Eq. (4) of the main text by noting that, for parabolic conduction and valence bands, $\langle \phi_{c/v\mathbf{k}} | \hat{e} \cdot \hat{p} | \phi_{c/v\mathbf{k}} \rangle = \pm k/m_{e/h}$ giving

$$\mathcal{M}_{mn}^{\text{exc}} = \left[ \frac{1}{m_h} + \frac{1}{m_e} \right] \sum_{\mathbf{k}} A_{\mathbf{k}}^{m*} A_{\mathbf{k}}^n \, \hat{e} \cdot \mathbf{k}. \qquad (9)$$

The sum in the RHS can be expressed as the expectation value of the dipole operator, $\sum_{\mathbf{k}} A_{\mathbf{k}}^{m*} A_{\mathbf{k}}^n \, \hat{e} \cdot \mathbf{k} \propto (E_{b,\text{exc}}^m - E_{b,\text{exc}}^n) \int d\mathbf{r} F^{m*}(r) \hat{e} \cdot \mathbf{r} F^n(r)$, which yields the typical hydrogen like selection rules $\Delta l = \pm 1$, with $l$ the angular momentum quantum number.

For the calculations on the MoS$_2$ polaritons we used an excitonic effective mass of 0.27

7

a.u. which is calculated from the $G_0W_0$ band structure. The Keldysh screened electron-hole interaction is calculated with a polarizability $\alpha = 13.5$ a.u.. A comparison of exciton binding energies obtained with the MW model and from the solution of the BSE is shown in Fig. S4. With the MW we are then able to calculate all the excitonic quantities required in

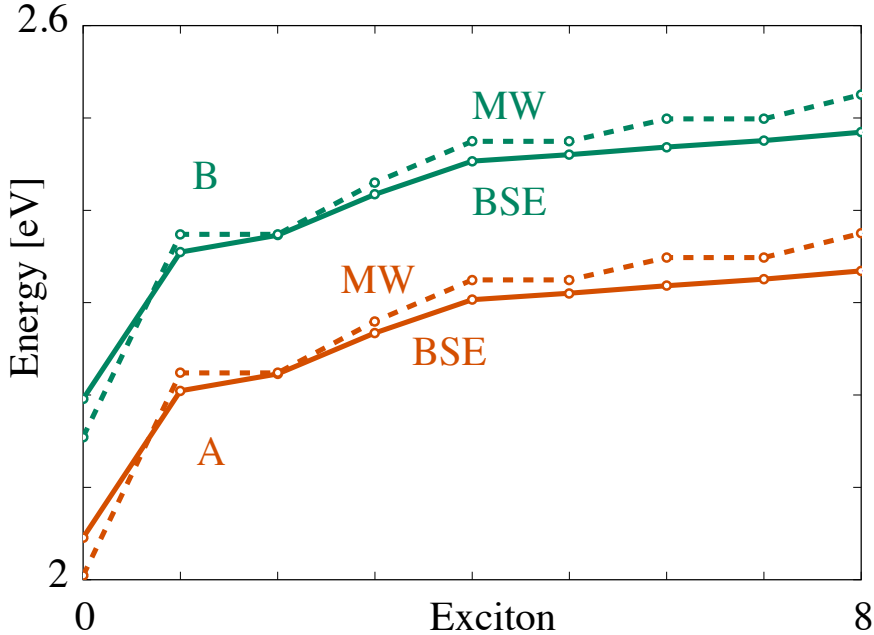

Figure S4: Comparison of excitons energies calculated using BSE and the Mott-Wannier (MW) model.

Eq. (2) of the main text and the same diagonalization procedure adopted in the QED-BSE can be performed to get the polaritonic states. Fig. S5 shows a side-by-side comparison of the polariton dispersion of $MoS_2$ obtained with the full QED-BSE approach and the QED-MW model. Small differences are apparent, that can be traced back to minor shifts in the exciton eigenvalues within the MW model, cf. Fig. S4 and differences in the momentum matrix elements. For a more detailed comparison, in Tab. S2 we reported values of the Rabi splittings and intensities of the response function for the "1s" A and "2s" A excitons within the QED-MW and QED-BSE methods.

We stress again that in the QED-MW, the exciton-polariton dispersion is now obtained at a much lower computational cost than for the QED-BSE case and at the same time a better

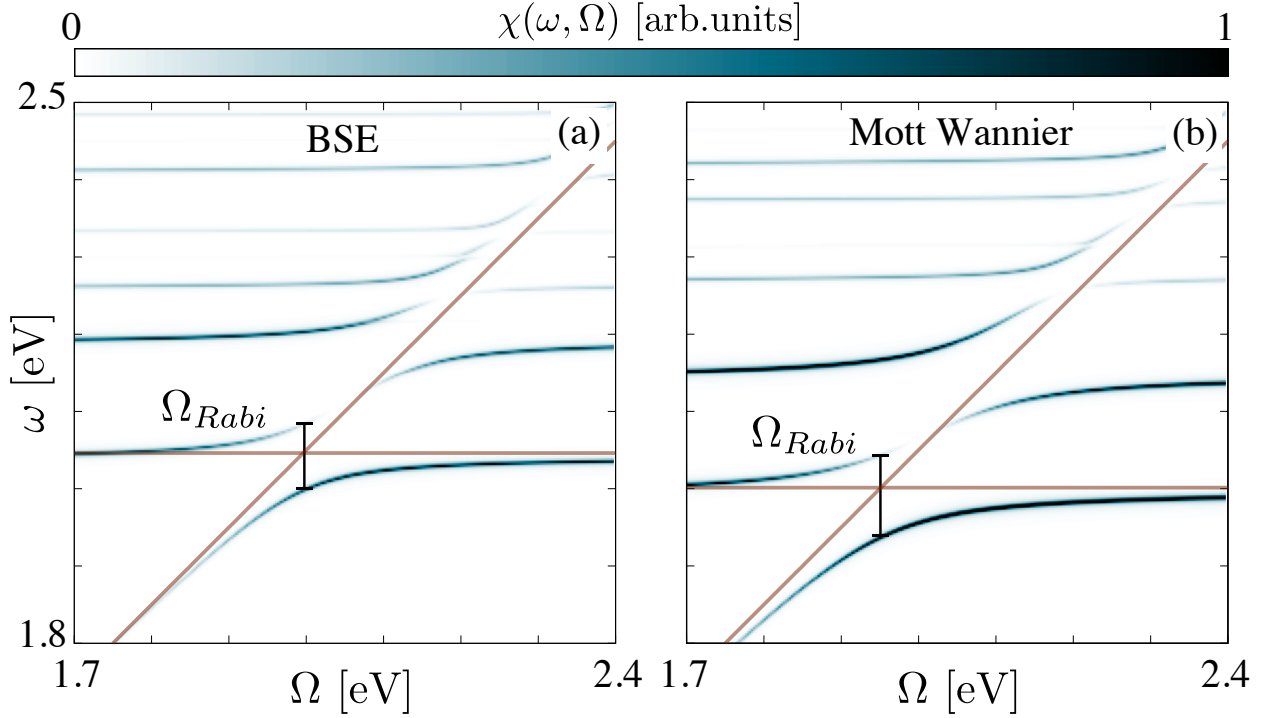

Figure S5: Comparison of QED-BSE and QED-MW for MoS$_2$ and with a coupling strength of $\tilde{A}_0 = 0.05$ a.u.

Table S2: QED-MW and QED-BSE comparison of Rabi splittings and response intensities for the 1s and 2s A excitons in MoS$_2$. The Rabi splitting are evaluated at the listed frequency $\Omega$ and the intensities are reported for the lower (LP) and upper (UP) polaritons with compatible arb.units.

| Excitation | $\Omega$ (eV) | Rabi Splitting (eV) | $\chi_{LP}$ (arb.units) | $\chi_{UP}$ (arb.units) |
|---|---|---|---|---|
| | | BSE | | |
| 1s A | 2.00 | 0.084 | 21.33 | 1.19 |
| 2s A | 2.16 | 0.041 | 0.75 | 7.45 |
| | | MW | | |
| 1s A | 1.95 | 0.106 | 31.39 | 2.94 |
| 2s A | 2.17 | 0.043 | 0.55 | 9.58 |

intuition on the photon mediated mixing of excitons can be built in terms of hydrogenic-like selection rules, according to Eqs. (5)-(9).

In the main text we used the MW model to access excitons in more complex materials, without increase in computational cost. In particular we studied the exciton-polariton dispersion in an MoS2 monolayer encapsulated in semi-infinite dielectric with a dielectric constant $\kappa$. For this particular stacking of materials, one can readily extend the QED-MW

for monolayers by including the effect of dielectric screening of the encapsulating material at three different levels: (i) the screened electron-hole Coulomb interaction is modified to[5]

$$W(\mathbf{r}, \kappa) = \frac{1}{4\alpha} \left[ H_0(\kappa x) - N_0(\kappa x) \right]_{x=r/2\pi\alpha} \qquad (10)$$

(ii) the bare photon dispersion in the dielectric medium changes to $\omega = \omega_c/\sqrt{(\kappa)}$ and (iii) the effective vector potential amplitude becomes $A_0 \to A_0/\kappa$.[6] The extra electronic screening due to the dielectric not only modifies the exciton binding energies but it also renormalizes the electronic gap. To estimate the electronic band gap variation as a function of $\kappa$, we take the standard assumption[7] that the change in the binding energy of the $1s$ exciton compensates the renormalization of the band gap, hence the condition $E_{\text{exc}}^{1s}(\kappa) = E_{\text{exc}}^{1s}(\kappa = 0)$ must hold. The results are shown in the main text in Fig. 4.

## Interlayer Excitons: Computational Details

In order to achieve an accurate first-principles description of the excitonic resonances in the $MoS_2$-$WS_2$ vdWH, we used the MW exciton model combined with the quantum electrostatic heterostructure (QEH) method in ref.[8] This method allows to account for dielectric screening effects on exciton binding energies and electronic band edges from ab-initio without restriction on lattice parameter matching between the two layers. We followed exactly the procedure used in ref.,[9] which for the sake of simplicity is summarized in the following. We first calculate GW band edges for each isolated monolayer and we then use the QEH-GW to include the effect of the dielectric screening induced by the adjacent layer and correct the electronic band edges position and band gap. In order to get the correct type II band alignment when compared to experiments,[10] norm-conserving PAW-setups are required.[11] We find a gap renormalization of $\sim 0.132$ eV for $MoS_2$ and $\sim 0.137$ eV for $WS_2$. Within the QEH we are then also able to calculate the screened interaction appearing in the MW model and calculate both interlayer and intralayer exciton binding energies. We find an interlayer

exciton binding energy of 0.28 eV and intralayer binding energy of 0.42 eV and 0.38 eV for $MoS_2$ and $WS_2$ respectively.

# References

(1) Mortensen, J. J.; Hansen, L. B.; Jacobsen, K. W. *Phys. Rev. B* **2005**, *71*, 035109.

(2) Enkovaara, J.; Rostgaard, C.; Mortensen, J. J.; Chen, J.; Dułak, M.; Ferrighi, L.; Gavnholt, J.; Glinsvad, C.; Haikola, V.; Hansen, H. A. et al. *J. Phys.: Cond. Mat.* **2010**, *22*, 253202.

(3) Haastrup, S.; Strange, M.; Pandey, M.; Deilmann, T.; Schmidt, P. S.; Hinsche, N. F.; Gjerding, M. N.; Torelli, D.; Larsen, P. M.; Riis-Jensen, A. C. et al. *2D Mater.* **2018**, *5*, 042002.

(4) Latini, S.; Olsen, T.; Thygesen, K. S. *Phys. Rev. B* **2015**, *92*, 245123.

(5) Massicotte, M.; Vialla, F.; Schmidt, P.; Lundeberg, M. B.; Latini, S.; Haastrup, S.; Danovich, M.; Davydovskaya, D.; Watanabe, K.; Taniguchi, T. et al. *Nat. Commun.* **2018**, *9*, 1633.

(6) Grosso, G.; Parravicini, G. *Solid State Physics*; Elsevier Science, 2000.

(7) Ugeda, M. M.; Bradley, A. J.; Shi, S.-F.; da Jornada, F. H.; Zhang, Y.; Qiu, D. Y.; Ruan, W.; Mo, S.-K.; Hussain, Z.; Shen, Z.-X. et al. *Nat. Mater.* **2014**, *13*, 1091.

(8) Andersen, K.; Latini, S.; Thygesen, K. S. *Nano Lett.* **2015**, *15*, 4616–4621.

(9) Latini, S.; Winther, K. T.; Olsen, T.; Thygesen, K. S. *Nano letters* **2017**, *17*, 938–945.

(10) Heo, H.; Sung, J. H.; Jin, G.; Ahn, J.-H.; Kim, K.; Lee, M.-J.; Cha, S.; Choi, H.; Jo, M.-H. *Advanced Materials* **2015**, *27*, 3803–3810.

(11) Klimeš, J.; Kaltak, M.; Kresse, G. *Physical Review B* **2014**, *90*, 075125.



\end{document}